\documentstyle[11pt]{article}
\newcommand{\ct}[1]{\cite{#1}}
\newcommand{\np}[1]{Nucl. Phys. {\bf B{#1}}}
\newcommand{\prv}[1]{Phys. Rev. {\bf {#1}}}
\newcommand{\prp}[1]{Phys. Rep. {\bf {#1}}}

\renewcommand{\theequation}{\thesection.\arabic{equation}}
\renewcommand{\thefootnote}{\fnsymbol{footnote}}

\newlength{\extraspace}
\newlength{\extraspaces}
\newcounter{dummy}

\newcommand{\baa}{
\addtocounter{equation}{1}
\setcounter{dummy}{\value{equation}}
\setcounter{equation}{0}
\renewcommand{\theequation}{\thesection.\arabic{dummy}\alph{equation}}
\begin{eqnarray}
\addtolength{\abovedisplayskip}{\extraspaces}
\addtolength{\belowdisplayskip}{\extraspaces}
\addtolength{\abovedisplayshortskip}{\extraspace}
\addtolength{\belowdisplayshortskip}{\extraspace}}

\newcommand{\eaa}{
\end{eqnarray}
\setcounter{equation}{\value{dummy}}
\renewcommand{\theequation}{\thesection.\arabic{equation}}}

\newcommand{\be}{\begin{equation}
\addtolength{\abovedisplayskip}{\extraspaces}
\addtolength{\belowdisplayskip}{\extraspaces}
\addtolength{\abovedisplayshortskip}{\extraspace}
\addtolength{\belowdisplayshortskip}{\extraspace}}
\newcommand{\ee}{\end{equation}}

\newcommand{\ba}{\begin{eqnarray}
\addtolength{\abovedisplayskip}{\extraspaces}
\addtolength{\belowdisplayskip}{\extraspaces}
\addtolength{\abovedisplayshortskip}{\extraspace}
\addtolength{\belowdisplayshortskip}{\extraspace}}
\newcommand{\ea}{\end{eqnarray}}

\newcommand{\bd}{\begin{displaymath}
\addtolength{\abovedisplayskip}{\extraspaces}
\addtolength{\belowdisplayskip}{\extraspaces}
\addtolength{\abovedisplayshortskip}{\extraspace}
\addtolength{\belowdisplayshortskip}{\extraspace}}
\newcommand{\ed}{\end{displaymath}}

\newcommand{\ban}{\begin{eqnarray*}
\addtolength{\abovedisplayskip}{\extraspaces}
\addtolength{\belowdisplayskip}{\extraspaces}
\addtolength{\abovedisplayshortskip}{\extraspace}
\addtolength{\belowdisplayshortskip}{\extraspace}}
\newcommand{\ean}{\end{eqnarray*}}

\newcommand{\nonu}{\nonumber}
\newcommand{\nt}{\noindent}

\newcommand{\hf}{{\displaystyle{1\over 2}}}                  
\newcommand{\hv}{{\displaystyle{1\over 4}}}                  
\newcommand{\hh}{{\displaystyle{1\over 16}}}                 
\newcommand{\kk}[2]{{\displaystyle{{#1}\over {#2}}}}

\newcommand{\lB}{\Bigg[}
\newcommand{\rB}{\Bigg]}
\newcommand{\lp}{\left(}                                  
\newcommand{\rp}{\right)}                                 
\newcommand{\la}{\langle}                                 
\newcommand{\ra}{\rangle}                                 

\newcommand{\ie}{{\it i.e., }}
\newcommand{\eg}{{\it e.g., }}

\newcommand{\id}{{\bf 1}}

\newcommand{\bb}{\beta}
\newcommand{\g}{\gamma}
\newcommand{\dd}{\delta}
\newcommand{\ep}{\epsilon}
\newcommand{\lbd}{\lambda}

\newcommand{\et}{\eta}

\newcommand{\s}{\sigma}
\newcommand{\gb}{\Gamma}
\newcommand{\db}{\Delta}

\newcommand{\ob}{\Omega}
\newcommand{\fb}{\Phi}
\newcommand{\fk}{\phi}
\newcommand{\ze}{\zeta}
\newcommand{\Om}{{\C{O}_m}}
\newcommand{\e}{\C{E}}
\newcommand{\C}[1]{{\cal {#1}}}
\newcommand{\B}[1]{\bar {#1}}
\newcommand{\spf}{[ \s ]}
\newcommand{\ef}{[ \C{E} ]}
\newcommand{\idf}{[ {\id} ]}
\newcommand{\tld}[1]{\tilde {{#1}}}
\newcommand{\si}{\psi}
\newcommand{\sib}{\bar{\psi}}
\font\cmss=cmss10
\font\cmsss=cmss10 at 7pt
\newcommand{\IZ}{\relax\ifmmode\mathchoice{\hbox{\cmss Z\\kern-.4em Z}}
{\hbox{\cmss Z\kern-.4em Z}}{\lower.9pt\hbox{\cmsss Z\kern-.4em Z}}
{\lower1.2pt\hbox{\cmsss Z\kern-.4emZ}}\else{\cmss Z\kern-.4emZ}\fi}
\newcommand{\II}{\relax{\rm I\kern-.18em I}}

\newcommand{\pr}[1]{\partial_{#1}}
\newcommand{\var}[1]{\frac{\partial}{\partial {#1}}}
\newcommand{\vars}[2]{\frac{\partial {#1}}{\partial {#2}}}
\newcommand{\ddl}{\frac{d}{d \ell}}
\newcommand{\had}[2]{\lim_{{#1} \rightarrow {#2}}~}
\newcommand{\ant}[2]{\int_{\scriptstyle{{#1}}}d^2{#2}~}
\newcommand{\Dant}[2]{\int_{\scriptstyle{{#1}}}d^D{#2}~}
\newcommand{\scr}[1]{{\scriptscriptstyle #1}}

\begin{document}
\input{epsf.tex}
\begin{titlepage}

\begin{flushright}
UCLA/93/TEP/49\\
hep-th/9312207
\end{flushright}

\vskip 1.5cm

\begin{center}
                {\bf{MANIFESTLY FINITE PERTURBATION THEORY\\
                         FOR THE SHORT-DISTANCE EXPANSION OF\\
                               CORRELATION FUNCTIONS IN\\
                       THE TWO DIMENSIONAL ISING MODEL}}
\end{center}
\vskip 1.25cm

\begin{center}
\normalsize{\bf{Bakhtiar Mikhak}}
\footnotesize{and}
\normalsize{\bf{Amir M. Zarkesh\footnote{Supported in part by D.O.E. contract
DE-FG03-91ER 40662 Task C.}}}
\end{center}

\begin{center}
{\it Physics Department\\
University of California, Los Angeles\\
Los Angeles, CA 90024, USA}
\end{center}
\vskip 1.25cm

\begin{center}
ABSTRACT
\end{center}

\indent
In the spirit of classic works of Wilson on the renormalization group and
operator product expansion, a new framework for the study of the theory space
of euclidean quantum field theories has been introduced. This formalism is
particularly useful for elucidating the structure of the short-distance
expansions of the $n$-point functions of a renormalizable quantum field theory
near a non-trivial fixed point. We review and apply this formalism in the study
of the scaling limit of the two dimensional massive Ising model.
Renormalization group analysis and operator product expansions determine all
the non-analytic mass dependence of the short-distance expansion of the
correlation functions. An extension of the first order variational formula to
higher orders provides a manifestly finite scheme for the perturbative
calculation of the operator product coefficients to any order in parameters. A
perturbative expansion of the correlation functions follows. We implement this
scheme for a systematic study of correlation functions involving two spin
operators. We show how the necessary non-trivial integrals can be calculated.
As two concrete examples we explicitly calculate the short-distance expansion
of the spin-spin correlation function to third order and the spin-spin-energy
density correlation function to first order in the mass. We also discuss the
applicability of our results to perturbations near other non-trivial fixed
points corresponding to other unitary minimal models.

\vskip 3cm
\noindent
December 93

\end{titlepage}
\renewcommand{\thefootnote}{\arabic{footnote}}
\setcounter{footnote}{0}


\setcounter{equation}{0}
\section{Introduction}
\vskip .25cm
\indent

The equivalence of statistical mechanics models and quantum euclidean field
theories with an ultraviolet cutoff has provided us with the exchange and
deeper understanding of very fruitful notions, such as the renormalization
group (RG) and spontaneous symmetry breaking, between the two disciplines. The
modern renormalization group language offers a beautiful qualitative
description of how the existence of second order (continuous) phase transitions
in a statistical mechanical system is relevant in defining a local quantum
field theory\ct{Wil2}. In particular the existence of a critical point and the
number of parameters one needs to fine-tune in order to reach the critical
point of a statistical model correspond to the existence of an ultraviolet
fixed point and the number of relevant operators of the continuum field theory,
respectively. In fact the exact solvability of statistical lattice models gives
us detailed quantitative information on the existence and proper definition of
the continuum limit of a quantum field theory and, in principle, allows for a
non-perturbative evaluation of all the correlation functions in the
theory\footnote{For an example in two dimensions see\ct{WMTB}.}.

In many situations where the lattice model is not exactly solvable, the  exact
solvability of the fixed point theory is expected to allow for perturbative
calculation of the correlation functions of the model near criticality. This is
the situation ordinarily encountered in field theory calculations where the
fixed point theory is trivial, \ie a free field theory. It is reasonable to ask
whether the same is true of {\em non-trivial\/} exactly solvable fixed point
theories. Fixed point theories are the fixed points of the renormalization
group action on the theory space of all quantum field theories. Due to the
complexity of theory space with respect to the RG flows, information about
theories not in the immediate neighborhood of a given ultraviolet fixed point
is accessible only non-perturbatively. It is believed that, in many physically
interesting cases, under the action of the renormalization group the theory
flows from an ultraviolet (UV) fixed point to an infrared (IR) fixed point in
the theory space.

It is very difficult to give a precise definition of the space of all quantum
field theories. In fact this definition is dependent on the field-theoretic
issues under consideration. For a better understanding of theory space it is
clearly important to have a perturbative scheme for studying off-critical
theories, in the vicinity of and away from their non-trivial fixed points,
given the information encoded in the fixed point theories only. When
off-critical theories are defined as perturbations of critical theories we
should in general consider perturbations by relevant, marginal, and also
irrelevant operators of the critical theory. Among these, the perturbations
which are integrable are of special interest. Comparison of perturbative
calculations in such models with exact results may shed light on how
non-perturbative information can be extracted from such analysis.

The greatest success in this direction has been in two
dimensions\ct{Zam1,Zam2}. This is thanks to the following two facts. First, in
two dimensions the scale invariance of a statistical system at its second order
phase transition point, combined with rotational and translational invariance,
implies full conformal invariance of the fixed point theory \ct{Pol1}. Second,
there are many completely integrable field theories which can be regarded as
perturbations of these critical models off criticality\ct{Zam2}. Therefore the
space of two dimensional field theories, particularly conformal and integrable
ones, provide the best theoretical playground for addressing many of the issues
mentioned above.

Two dimensional conformal field theory, due to its central role in string
theory \ct{String} and possible classification of all two dimensional critical
phenomena\ct{BPZ}, has been studied extensively in the past several
years\footnote{See articles in ref.\ct{CFT} also.}. Conformal field theories,
besides describing critical statistical systems, give ultraviolet fixed points
of the associated renormalizable quantum field theories. Though conformal
invariance is enough to determine all correlation functions of the fixed point
theories, these theories are by no means trivial. Unitary minimal models of
BPZ\ct{BPZ}, the simplest of which correspond to the critical and the
tri-critical Ising models, are all examples of such theories. Even though these
theories admit free field representations, they are still non-trivial, in light
of the fact that conformal symmetries only admit coordinate space
realizations\footnote{For these models the standard Feynman techniques are not
applicable.}.

Certain perturbations of minimal models are known to be completely integrable.
The simplest example of this is the thermal perturbation of the two dimensional
Ising model. A much less trivial example is the RG flow between tri-critical
and critical Ising models. It is believed that a suitable integrable
perturbation of the tri-critical Ising model has the critical Ising model as
its IR fixed point\ct{Zam1,flow}. Such flows, aside from being physically
interesting in their own right, provide useful toy models for higher
dimensional low energy effective field theories of completely solvable high
energy theories. These provide ample reason for a need for a calculational
scheme which naturally takes advantage of the existence of {\em non-trivial}
yet solvable fixed point theories. Even though there has been some success in
doing perturbative calculations near non-trivial fixed points of specific
models up to low orders, they have not lead to a systematic calculational
scheme up to now.

Recently, expanding upon ideas originally put forth by Wilson \ct{Wil2,Wil1},
and expanded upon by Wegner\ct{Weg1,Weg2}, a new framework for the analysis of
the theory space of all renormalizable $D$ dimensional euclidean quantum field
theories has been introduced (\ct{Son1}-\ct{Son8}). Though this formalism is in
principle non-perturbative, it provides a concrete program for perturbative
analysis and calculation of the short-distance expansion of the correlation
functions of a renormalizable field theory around a {\em non-trivial\/} fixed
point. The first step in this program is to make a particular choice of
coordinates for the infinite dimensional theory space such that all but a
finite number of coordinates vanish for renormalizable theories. With this
choice we can analyze the structure of the renormalization group equations by
counting the scale dimensions of the coordinates\ct{Son1,Son3}.

The next step is to combine the renormalization group with the operator product
expansion (OPE) and determine all non-analytic dependence on the renormalized
parameters in the short-distance expansion of the $n$-point correlation
functions \ct{Weg2}. At this point, all that remains for us to calculate are
the RG invariant functions appearing in the solution of the RG
equations\ct{Son1}. Their evaluation requires an explicit calculation based on
a concrete regularization scheme, which was provided by the introduction of a
variational formula for correlation functions in coordinate space\ct{Son4}. In
this formalism the operator product expansion is introduced and incorporated in
the definition of the variational formula in order to regulate all the
ultraviolet divergences. It is not necessary to establish the validity of the
OPE by extensive calculations\ct{Wil1} as it is already built into the
formalism (\ie OPE is assumed).

The variational formula for the correlation functions realizes the derivative
of correlation functions with respect to a parameter in terms of an insertion
of a conjugate operator. The notion of conjugacy is the same as the one used in
thermodynamics, in the sense that the expectation value of any such operator is
defined to be the derivative of the free energy density of the system. The
existence of such operators is an assumption. The variational formula is
formulated in coordinate space, and involves divergent subtractions due to
short-distance singularities of the renormalized theory. We also need to
introduce finite counterterms to compensate for the arbitrariness of the
subtraction scheme. The finite counterterms can be interpreted geometrically as
connections in the theory space\ct{Son4,Son8}.

The variational formula is very hard to derive from first principles and we can
only demand its consistency with RG and the usual properties of derivatives.
Requiring consistency with the RG equations leads to a relationship between the
anomalous dimensions of the composite operators and the unintegrable part of
the operator product expansion\ct{Son4}. The curvature of the above connection
appears in this relationship naturally. The commutativity of mixed higher order
derivatives (Maxwell relations) leads to an expression for the curvature of the
connection in terms of a double integral over a finite domain. This provides us
with a very natural setting for studying issues relating to general covariance
and geometry in theory space\ct{Son7,Son8}. Upon applying the variational
formula for correlation functions to the OPE we arrive at a variational formula
for the operator product coefficients\ct{Son5}. Repeated application of these
variational formulas allows for a perturbative calculation of the correlation
functions.

The large number of exact results in two dimensional conformal and integrable
field theories motivates the application of the above formalism to the study of
the space of two dimensional quantum field theories. Since the above framework
and its full conceptual and calculational power is not yet popular we decided
to present a comprehensive discussion of the systematic calculational scheme
which is contained in this formalism. In particular we illustrate the
usefulness of this scheme in analyzing a historically significant model, the
two dimensional Ising model. This model corresponds to the best understood
non-trivial fixed point in the space of two dimensional field theories. It was
solved exactly nearly half a century ago\footnote{For the Onsager solution of
the Ising model, see ref. \cite{SML} and references therein.}, and has served
as an important toy model for the study of existence and definition of
continuum field theories starting from lattice models\ct{WMTB}. The correlation
functions of the model are known exactly, thanks to its complete integrability
at and away from the fixed point\ct{WMTB,BPZ,DSZ}. The short- and long-distance
expansions of the correlation functions encode the operator content of the UV
and IR fixed point theories. It is desirable to understand the structure of
these expansions field-theoretically. Therefore, our interest in this model is
twofold. First, since the presence of the spin field in the Ising model does
not allow us to apply the traditional Feynman approach to perturbative
calculations in field theory, the use of the new formalism is essential.
Second, we can check the results of our calculations againt the available exact
results. The crucial point is that this method is in principle equally powerful
for many other interesting models for which exact solutions are not known.

The purpose of the present paper is to discuss the formalism sketched above in
a general renormalizable field theory with a UV fixed point, and to extend the
proof of infrared finiteness of the variational formula for the operator
product coefficients to any order. This essentially amounts to the consistency
of the variational formula with the analyticity of the OPE coefficients and
leads  to a manifestly finite variational formula for general operator product
coefficients\footnote{These are the coefficients appearing in the expansion of
correlation functions in terms of expectation value of single operators.}. We
therefore arrive at a systematic method for the perturbative calculation of OPE
coefficients, and consequently correlation functions, in the vicinity of any
non-trivial ultraviolet fixed point. As an application of our results we
calculate the spin-spin and the spin-spin-energy density correlation functions
in the massive Ising model perturbatively.

This paper not only provides the conceptual and calculational details of our
previous work\ct{MZ1}, but also extends the calculational framework of ref.
\ct{Son4,Son5} to higher orders. Our manifestly finite calculational scheme is
a product of a complete examination of the infrared structure of the higher
order variational formulas, and demanding consistency with locality principles.
Calculation of non-trivial two- and three-point functions in the two
dimensional Ising model within this scheme are explicit examples of the
practicality of this formalism in addition to its conceptual clarity. These are
the first field-theoretic calculations near a non-trivial fixed point
\footnote{Dotsenko\ct{Dot} has calculated the second order correction to the
spin-spin correlation function, but, as he points out, the generalization of
his analytical continuation method to other models seems formidable.} and
should, in principle, be generalizable to many other two dimensional models.
The main obstacle for carrying out the same analysis for other models is
finding a systematic way of calculating the necessary {\em finite} integrals,
which are quite complicated in general. However, our success in finding a
systematic technique for calculating the integrals for the Ising model
correlation functions gives us hope that the same should be possible at least
for other unitary minimal models.

The organization of the paper is as follows. In section 2 we review the RG
analysis of refs. \ct{Son1}-\ct{Son3} in general and in the two dimensional
Ising model. In section 3 we review the variational formulas \ct{Son4,Son5} and
discuss their ultraviolet and infrared divergence structures in detail. We also
introduce a variational formula for general operator product expansion
coefficients in this section. The consistency of the variational formula with
the differentiability of the OPE coefficients  to a given order is shown to be
equivalent to the IR finiteness of the variational formula to that order.
Hence, the analyticity of the OPE coefficients implies that the variational
formula is manifestly free of divergences to any order. In section 4 we analyze
the UV structure of the variational formulas for the Ising model. Here we fix
our choice of the parameters used in the remainder of the paper. In section 5
we apply the results of section 3 to the Ising model. We describe the structure
of the $n^{\underline{\rm th}}$ order correction to the scaling behavior of the
short-distance expansion of the spin-spin correlation function, and explain how
the required integrals can be performed. For completeness, the details of the
explicit calculation to third order\ct{MZ1} are presented in appendix B. In
section 6, as another non-trivial application of our results, we discuss the
calculation of the spin-spin-energy density correlation function to first
order. In section 7 we present our conclusions. In appendix A some relevant
facts about the operator content and the correlation functions of the critical
Ising model are presented.
\vskip .75cm
\pagebreak[3]


\setcounter{equation}{0}
\section{Renormalization Group Analysis}
\vskip .25cm

\subsection{General Theory}
\vskip .2cm
\nopagebreak

\noindent \underline{The Action of the Renormalization Group on the Theory
Space}
\nopagebreak
\indent

In the Wilsonian approach to renormalization we consider the {\em theory space}
($TS$) of all regularized quantum field theories in $D$ dimensional euclidean
space\ct{Wil2}. The theories in $TS$ are regularized with an ultraviolet cutoff
$a$ which is assumed to be invariant under both rotations and translations. The
cutoff may be taken to be a lattice unit as long as the continuum theory
obtained in the limit of vanishing lattice spacing is rotationally and
translationally invariant. Let $\C{H}$ denote a point in the theory space
representing the action of a $D$ dimensional euclidean quantum field
theory\footnote{Equivalently $\C{H}$ can be regarded as the hamiltonian of a
$D$ dimensional statistical system.}.

The action of the renormalization group (RG), denoted by $\C{R}(\ell)$, is
parametrized by a real variable $\ell$ and is defined as an integration over
fluctuations between $a$ and $ae^{-\ell}$ followed by a dilation by a factor
$e^{-\ell}$. Hence, the distance $r$, measured in units of the cutoff, is
transformed to $re^{-\ell}$, while the renormalization point $r=1$ is fixed.
Note that RG acts toward the infrared and differs from the standard definition
by rescaling. We generally assume the existence of fixed points in $TS$ with
respect to the RG transformations. Let $\C{H^\ast}$ denote one such point
defined with the property

\be
\C{R}(\ell) \C{H^\ast} = \C{H^\ast} \;\;\;\;\;\; (\forall \ell \in \bf{\rm R}).
\ee

\nt At the fixed point scale invariance implies

\be \label{scaling}
\la \B{\fk}_{i_1}(\lbd r_1) \cdots \B{\fk}_{i_n}(\lbd r_n) \ra_{\C{H^\ast}} =
\frac{1}{\lbd^{\scriptstyle (x_{i_1}+\cdots+x_{i_n})}} \la \B{\fk}_{i_1}(r_1)
\cdots \B{\fk}_{i_n}(r_n) \ra_{\C{H^\ast}},
\ee

\nt and we can assume that the operators $\B{\fk}_i$ have non-negative scale
dimensions $x_i$, and are called relevant, marginal, or irrelevant if $x_i$ are
smaller than, equal to, or greater than $D$, respectively.

The hamiltonian of a theory in the neighborhood of $\C{H^\ast}$ is obtained by
adding a linear combination of local scalar\footnote{Thanks to the assumption
of rotational invariance we need to only consider scalar operators.} operators
to the fixed point hamiltonian. Let $\{ \B{\fk}_i \}$ denote a complete basis
of local scalar operators which are infinite in number. The generalized mass
parameters $\{\mu^i\}$, introduced by Wilson, are the coefficients of these
local scalar operators and provide local coordinates for the $TS$. Hence, $TS$
is infinite dimensional.

Generally, starting from any hamiltonian $\C{H}$ in the neighborhood of
$\C{H^\ast}$, RG transformations $\C{R}(\ell)$ generate a trajectory of
hamiltonians $\C{R}(\ell)\C{H}$ parametrized by $\ell$. The range of $\ell$ can
be extended in both directions as long as the trajectory stays within the
theory space. Not all trajectories can be extended without limit. A physically
interesting subspace of $TS$ is the space of renormalized trajectories, denoted
by $S(\infty)$, consisting of all trajectories which can be traced back to
$\ell = - \infty$ without going out of the theory space.\footnote{This is
equivalent to the assumption of the existence of a UV fixed point at the origin
of renormalized trajectories.} Renormalizable theories are typically
characterized by a finite number of renormalized parameters and we should
therefore be able to parametrize $S(\infty)$ by a finite number of parameters
forming a finite dimensional subspace of the theory space, accessible
asymptotically in the ${\ell}\rightarrow{\infty}$ limit of the RG flows. Let us
explain these points more carefully\ct{Wil2,Par}.

Consider the lattice models parametrized by couplings $\kappa^i$ \footnote{This
is one particular choice for Wilson's generalized mass parameters.} where $i\in
\C{I}$ with $\C{I}$ an (in general infinite) index set. Assume that the model
undergoes a second order phase transition at $\kappa^{(c)i}$ parametrizing the
critical hamiltonian $\C{H}_{lattice}^{(c)}$. At the critical point all
parameters have well-defined scale dimensions, $y_i = D - x_i$, and we assume
that only a finite number of them have positive scale dimensions. Starting from
an almost critical theory, under RG transformations the effect of the couplings
with negative scale dimensions gets washed out. Therefore in the infrared limit
the action of the renormalization group singles out a finite dimensional
subspace of the theory space corresponding to a renormalized trajectory, \ie
{\em universality}. The dimension of this subspace is equal to the number of
relevant operators conjugate to couplings with positive scale dimensions.

When the continuum theory exists, with the proper choice of coordinates (which
will be described shortly) $\C{H}_{lattice}^{(c)}$ flows to the origin of this
subspace corresponding to $\C{H^\ast}$ which is the UV fixed point of a
renormalized field theory originating from it. In practice after reaching the
renormalized trajectory in the infrared limit we can trace back to $\C{H^\ast}$
along the renormalized trajectory\footnote{Since proving the existence of UV
fixed points is in general very difficult, we assume it. In principle these
fixed point theories are fully interactive.}. Therefore these theories are
defined up to infinitesimal distances and on the trajectory the parameter $a$
plays the role of a renormalization point. In the process of defining the
continuum limit in this way we are led to considering a finite number of
renormalized parameters, $g^i$, characterizing this trajectory, which are
analytic functions of the couplings $\kappa^i$. Therefore the goal would be to
find a particular choice of renormalized parameters, which allows for the study
of the non-analytic dependence on the generalized mass parameters in the
short-distance expansion of the correlation functions. In the next subsection
we will see how this happens for the Ising model.

While Wilson's generalized mass parameters provide local coordinates for the
infinite dimensional theory space, they are not necessarily the most suitable
choice for a smooth parametrization of the renormalized trajectories. First,
since we only need a finite number of renormalized parameters to parameterize
the renormalized trajectories, we would expect that there must exist a new set
of coordinates for $TS$ such that all but a finite number of them vanish on
$S(\infty)$. Second, it would be desirable to be able to analyze the structure
of the RG equations for these new coordinates by counting scale dimensions
only. The advantage of these new coordinates would be that their scale
dimensions are additively conserved under RG flows. This, in addition to the
analyticity requirement of the RG equations implied by the locality of the
theory, restricts the form of the RG equations to contain mixing with
parameters of lower or equal scale dimensions only.

Based on Wegner's scaling fields \ct{Weg1}, it has been shown that the analytic
transformation of interest to the new coordinates $g^i (i=1 \cdots N)$ exists
\ct{Son1}. Under this change of coordinates, which is assumed to be analytic,
our basis of local operators transforms like a vector. Denote the new basis of
composite operators by $\{\fk_i\}_{g}$.  In terms of these new coordinates, the
operators $\fk_i$ have well-defined short-distance singularities only for the
theories on $S(\infty)$. This allows for a systematic determination of the
corrections to the scaling behavior of the short-distance expansion of the
correlation functions. Note that if we consider general theories not restricted
to $S(\infty)$, the correlation functions contain arbitrarily high
singularities, even in terms of the new coordinates.

The coordinates of the theory space can be chosen such that the action of the
renormalization group on the renormalized trajectories is

\be \label{beta}
\ddl g^i = \bb^i (g) \equiv y_{i}g^i + \sum_{I; y_{i_1},\cdots,y_{i_n} \geq 0
\atop y_I = y_i} \frac{1}{n!} \bb_{i,I}g^I,
\ee

\nt where $I$ stands for $i_1 \cdots i_n$, $y_I = y_{i_1} +  \cdots + y_{i_n}$
and $g^I \equiv g^{i_1}\cdots g^{i_n}$. The coefficients $\bb_{i,I}$ are
independent of $g$. Note that $g=0$ is kept invariant under (\ref{beta}). In
ref. \ct{Son1} it was shown that with this choice of coordinates the short
distance limit reduces to the limit of $g^i \rightarrow 0$.
\vskip .5cm
\pagebreak[2]


\noindent \underline{Vector Bundle of Composite Operators}
\indent

Each point $g$ in the space of renormalizable theories is a renormalized field
theory which has an infinite number of linearly independent composite fields.
$\{\fk_{a}\}_{g}$ denotes a complete basis of composite fields at $g$. The
space of composite fields is a linear space, and therefore forms an infinite
dimensional vector bundle over $S(\infty)$\ct{Son8}. The choice of the local
basis is by no means unique, and two bases $\{\phi_{a}\}_{g}$ and
$\{{\phi^{\prime}_{a}}\}_{g}$ are related by an invertible infinite dimensional
matrix $N(g)$ via

\be \label{N}
{\fk^{\prime}}_{a} = {(N(g))_{a}}^{b} \fk_{b}.
\ee

\noindent The physical quantities, the correlation functions of $n$ composite
fields $\la {\phi}_{a_{1}}(r_{1}) \cdots {\phi}_{a_{n}}(r_{n}) \ra_{g}$, form
rank-n tensors on the theory space.

Among the infinite number of composite fields, $N$ composite scalar fields
stand out; they are the fields $\C{O}_{i}$ {\em conjugate} to the parameters
$g^{i}$. The notion of conjugacy should be familiar from thermodynamics. The
derivative of the free energy density $F(g)$ with respect to $g^i$ is given by
the expectation value of $\C{O}_{i}$\footnote{In order to make the notion of
conjugacy in field theories precise, we also need to specify how the derivative
of a correlation function is realized. The variational formula (VF) in section
3 is one such realization.}:

\be \label{con}
\la \C{O}_{i} \ra_{g} = \var{g^i} F(g).
\ee

\noindent The RG action (\ref{beta}) combined with canonical scaling of the
free energy density

\be \label{RGF}
\ddl F(g) = D F(g),
\ee

\noindent and (\ref{con}) induces an RG equation for composite operators
$\C{O}_{i}$:

\be \label{RG1}
\ddl \C{O}_{i} = D \C{O}_{i} - \vars{\bb^j}{g^i} \C{O}_{j}.
\ee

The particular choice of parameters \{$g^i$\} restricts the form of RG
equations for the general composite fields; scale dimensions are additively
conserved, and under RG any field can only mix with operators of equal or
smaller scale dimensions, with mixing coefficients analytic in the parameters.
Therefore, introducing an infinite column vector $\Phi$ of composite operators,
we can write

\be \label{RG1c}
\ddl \fb = \gb(g) \fb,
\ee

\noindent where the matrix

\be \label{anom}
{\gb_{a}}^{b}(g) \equiv {x_{a}\dd_{a}}^{b} + O(g)
\ee

\noindent gives the scale dimensions $x_{a}$ and the anomalous dimensions of
renormalized fields. The precise meaning of the above RG equations is given by
the following RG equations for the correlation functions:

\ba \label{RGn}
\ddl \la {\phi}_{a_{1}}(r_{1}) \cdots {\phi}_{a_{n}}(r_{n}) \ra_{g} &\equiv&
\lp -\sum_{k=1}^{n} r_{k,\mu}\var{r_{k,\mu}} + \bb^i \var{g^i} \rp \la
{\phi}_{a_{1}}(r_{1}) \cdots {\phi}_{a_{n}}(r_{n}) \ra_{g} \nonumber\\
&=& \sum_{k=1}^{n} \gb_{a_k}{}^b (g) \la {\phi}_{a_{1}}(r_{1}) \cdots
{\phi}_{b} (r_{k}) \cdots  {\phi}_{a_{n}}(r_{n}) \ra_{g}.
\ea
\vskip .5cm
\pagebreak[2]


\noindent \underline{Operator Product Expansion}
\nopagebreak
\indent

The operator product expansion (OPE) plays a central role in the analysis of
the short-distance expansion of correlation functions\ct{Wil1}. Wegner combined
OPE with the RG equations to ascribe all non-analytic dependence on the scaling
fields to the vacuum expectation values of single composite operators\ct{Weg2}.
When $r$ defined by $r^2 \equiv \sum_{i<j} (r_{i}-r_{j})^2$, is small compared
with the correlation length, we can write

\be \label{OPEn}
{\phi}_{a_{1}}(r_{1}) \cdots {\phi}_{a_{n}}(r_{n}) = {C_{a_{1} \cdots
a_{n}}}^b(r_{1}, \cdots, r_{n};g) \fk_{b}(0),
\ee

\noindent where ${C_{a_{1} \cdots a_{n}}}^b$ is invariant under translation.
The validity of this equation as an operator equation requires the sum over $b$
to run over all local operators with or without vanishing spins. This is
understood as an asymptotic expansion of the correlation functions

\be \label{OPEc}
\la {\phi}_{a_{1}}(r_{1}) \cdots {\phi}_{a_{n}}(r_{n}){\phi}_{d_{1}}(\xi_{1})
\cdots {\phi}_{d_{k}}(\xi_{k}) \ra_{g} =
{C_{a_{1} \cdots a_{n}}}^b (r_{1} \cdots r_{n};g) \;\la
\fk_{b}(0){\phi}_{d_{1}}(\xi_{1}) \cdots {\phi}_{d_{k}}(\xi_{k})\ra_{g}.
\ee

The fundamental assumption for the operator product expansion is that the
coefficients ${C_{a_{1} \cdots a_{n}}}^b(g)$ depend on the parameters only
analytically. This is equivalent to the local nature of the operator product
expansion: short-distance physics alone guarantees (\ref{OPEc}) to be valid.

Using the RG equation (\ref{RG1c}) we can derive the RG equation for the
coefficient functions

\be \label{RGcoeff}
\ddl {C_{a_{1} \cdots a_{n}}}^b(r_1,\cdots,r_n;g) = -\;{C_{a_{1} \cdots
a_{n}}}^d {\gb_{d}}^b(g)
+ \sum_{i=1}^{n} {\gb_{a_i}}^d(g) {C_{a_{1} \cdots a_{i-1}da_{i+1} \cdots
a_{n}}}^b,
\ee

\noindent where the matrix $\gb(g)$ is given by (\ref{anom}).
\vskip .5cm
\pagebreak[2]


\noindent \underline{Solution of RG Equations}
\nopagebreak
\indent

In order to solve RG equations (\ref{RGn}) and (\ref{RGcoeff}) we define the
running parameters $\B{g}(\ell;g)$ as the solutions of RG equations

\be \label{run}
\var{\ell} \B{g}^i(\ell;g) = \bb^i(\B{g}(\ell;g)),
\ee

\noindent with the initial conditions $\B{g}^i(0;g) = g^i$. We will suppress
these initial conditions from now on, and we will denote $\B{g}^i(\ell;g)$
simply by $\B{g}^i(\ell)$. We can then construct $N\;$ RG invariant quantities,
$\B{g}^i(\ln r)$. The RG invariance follows from the fact that the dependence
on the change of the coordinate $r$ under the RG is cancelled by the dependence
on the change of initial parameters $g$ under the RG.

We also need to introduce the matrix $G(r;g)$ that satisfies

\be \label{green}
\ddl G(r;g) = \gb(g) G(r;g),
\ee

\noindent with the initial condition $G(1;g) = \id$. The matrix $G(r;g)$ is
easily calculable for any field theory with at most one marginal parameter.
Using this matrix the solution to equation (\ref{RGcoeff}) is

\be \label{Rgcoeffs}
{C_{a_1 \cdots a_n}}^d(r_1,\cdots, r_n;g) = {G_{a_1}}^{b_1}(r;g) \cdots
{G_{a_n}}^{b_n}(r;g) {H_{b_1 \cdots b_n}}^e\Big(\frac{r_i-r_j}{r};\B{g}(\ln
r)\Big) {(G^{-1})_{e}}^d(r;g),
\ee

\noindent where

\be \label{genH}
{H_{a_1 \cdots a_n}}^d\Big(\frac{r_i-r_j}{r};g\Big) = {C_{a_1 \cdots
a_n}}^d\Big(\frac{r_1}{r},\cdots,\frac{r_n}{r};g\Big)
\ee

\noindent is an analytic function of $g$, due to the analyticity assumptions
for ${C_{a_{1} \cdots a_{n}}}^d(g)$. Therefore the solution of (\ref{RGn}) is

\be \label{RGns}
\la {\phi}_{a_{1}}(r_{1}) \cdots {\phi}_{a_{n}}(r_{n}) \ra_g =
{G_{a_1}}^{b_1}(r;g) \cdots {G_{a_n}}^{b_n}(r;g)
{H_{b_1 \cdots b_n}}^d\Big(\frac{r_i-r_j}{r};\B{g}(\ln r)\Big) \la \fk_d
\ra_{\B{g}(\ln r)},
\ee

\nt and all the non-analytic dependence on $g$'s in the correlation functions
comes from the expectation values of single operators (\ref{con}) through the
non-analyticity of the free energy density.

Since the short-distance limit on renormalized trajectories is equivalent to
the limit $g \rightarrow 0$, we obtain the short distance behavior \ct{Son1}

\be \label{sdl}
\la {\phi}_{a_{1}}(r_{1}) \cdots {\phi}_{a_{n}}(r_{n}) \ra_g \rightarrow  \la
{\fk}_{a_1}(r_1) \cdots {\fk}_{a_n}(r_n) \ra_{\C{H^\ast}}.
\ee

\nt which is consistent with our previous comment that the operators $\fk_i$
have well-defined short distance singularities only for the theories on
$S(\infty)$.
\vskip .25cm
\pagebreak[2]


\subsection{The Two Dimensional Ising Model}
\vskip .2cm
\nopagebreak
\indent

The Ising model is defined on a periodic lattice by the hamiltonian

\be
\C{H}_{lattice} \equiv -K \sum_{m,n=0}^{N} \{ \s_{m,n}\s_{m+1,n} +
\s_{m,n}\s_{m,n+1}\},
\ee

\noindent where $\s_{m,n} = \pm 1$. This theory undergoes a second order phase
transition at $K = K_c$ given by

\be
{\rm sinh}\;(2K_c) = 1.
\ee

The modern renormalization group language offers a qualitative description of
continuous phase transitions. Let us describe the phase transition in the Ising
model in this language. $\C{H}_{lattice}$ defines a one parameter family of
theories in the theory space. The action of the renormalization group, in the
infrared limit, singles out a renormalized trajectory in the theory space such
that theories with $K<K_c$ and $K>K_c$ flow to two different tails of this
line. The critical theory $\C{H}_{lattice}^{(c)}$, for $K = K_c$, flows to
$\C{H^\ast}$ (the origin of the coordinates on the renormalized trajectories)
along a critical line. On the renormalized trajectory, the vicinity of the
origin characterizes the short-distance behavior of the scaling limit of the
lattice theory. This limit is defined by letting ${\ep \equiv 4(K_c-K)}
\rightarrow {0}$, but keeping the product $t = \ep R$ ($R$ is the distance in
lattice units) finite. The renormalized mass $m$ should be proportional to
$\ep$ up to first order. The complete dependence of $m$ on $\ep$ is fixed by
demanding a simple RG equation for $m$.

In a statistical mechanical system near its second order phase transition
point, where the correlation length diverges, information about the details of
the lattice is washed out. If the above scaling limit exists we have a
well-defined quantum field theory all the way up to zero distance. This fact is
usually stated as the existence of an unstable UV fixed point which can be
reached backward along the renormalized trajectory, using the RG equations for
the renormalized parameters. We recover rotational invariance in the continuum
limit, and the theory is equivalent to a theory of a free massive Majorana
fermion. The theory is parametrized by the cosmological constant $g_{\id}$ and
the mass $m$, which are conjugate to the operators $\id$ and $\C{O}_m$
(corresponding to the energy density operator $i\B{\psi}\psi$), respectively.
These parameters satisfy the following RG equations

\be \label{RGg}
\ddl g_{\id} = 2 g_{\id} + \hf \bb_{\id} m^2
\ee
\be \label{RGm}
\ddl m = m.
\ee

\noindent There are two RG invariant quantities

\be
\B{g}_{\id}(\ln r) = r^2 g_{\id} + \hf \bb_{\id} m^2 r^2 \ln r
\ee
\be
\B{m}(\ln r) = mr.
\ee

In this case (\ref{RG1}) reduces to

\be
\ddl \C{O}_m = \C{O}_m - \bb_{\id} m \id.
\ee

\nt Taking the expectation value of both sides and using (\ref{RGm}) we have

\be \label{Om}
\la \C{O}_m \ra_{m} = -\bb_{\id} m \ln (mc),
\ee

\nt where c is a constant of integration. Equations (\ref{Om}) and (\ref{con})
give

\be \label{Fsol}
F(g_{\id},m) = g_{\id} - \hf\bb_{\id}m^2\ln(mc) + \hv\bb_{\id}m^2.
\ee

\noindent This was one of the main results of ref.\ct{Son2}.

Since $g_{\id}$ only contributes an additive constant to the free energy
density, none of the correlation functions of local operators depend on it. In
this case, we can effectively regard $S(\infty)$ as a line parametrized by a
real parameter $m$. This is the line corresponding to the theory of a free
massive Majorana fermion mentioned above. Therefore from now on all of our
physical quantities will depend on $m$ only.

The RG analysis of one point functions of more general composite operators in
the Ising model is quite straightforward. The first reason for this, as shown
in appendix A, is that the couplings to all the members of class $\s$ can be
turned off, due to the locality of the RG transformations and the special OPE
structure of the operators of the critical Ising model (see \ref{IsOPE}). The
second reason is that, due to the additive conservation of the scale dimension
of operators under RG transformations, the matrix $\gb(m)$ has the following
structure

\baa \label{anom2}
{\gb_{a}}^b(m) &=& x_a {\dd_{a}}^{b} + {\g_{a}}^{b} m^{x_a-x_b} ~~~~~~~~~~{\rm
iff}~~x_a-x_b\in \sf{Z \hspace{-1.65mm} Z}^{+} \\
&=& x_a {\dd_{a}}^{b} ~~~~~~~~~~~~~~~~~~~~~~~~~~\;\;{\rm otherwise.}
\eaa

\nt Note that (\ref{anom2}) is particularly simple since there is no marginal
(\ie dimension 2) scalar operator in this theory. Using (\ref{RGm}) and
(\ref{anom2}) we can write (\ref{RG1c}) as

\be
\ddl[m^{-X}\la \fb \ra_m] = \g [m^{-X}\la \fb \ra_m],
\ee

\nt where we have introduced the infinite dimensional diagonal matrix $X \equiv
diag[x_i]$.

Now since we are considering a unitary model, the lowest dimension operator is
the identity operator with scale dimension zero. Arrange the column vector
$\fb$ such that  operators with higher scale dimensions are lower in the
column. With this choice the matrix $\g$ is lower triangular. Therefore upon
repeated differentiation of the above we get a non-zero result

\be
\frac{d^n}{d\ell^n} [m^{-x_a} \la \C{\fk}_a \ra_m] = {\g_a}^{b_1}
{\g_{b_1}}^{b_2} \cdots {\g_{b_{n-1}}}^{b_n} [m^{-x_{b_n}} \la \C{\fk}_{b_n}
\ra_m],
\ee

\nt if and only if $a>b_1>\cdots>b_n$. We therefore see that the case
$\C{\fk}_{b_n}=\id$ corresponds to the largest value $n$ can have. Using
equation (\ref{RGm}) we see that $n$ is the highest power of $\ln (mc)$ and is
equal to the number of scale dimensions smaller than $x_a$, where at least one
non-derivative scalar composite operator exists. We therefore have

\be \label{log}
\la \C{\fk}_a \ra_m = m^{x_a} \frac{(-1)^n}{n!} {\g_a}^{b_1} {\g_{b_1}}^{b_2}
\cdots {\g_{b_{n-1}}}^{b_n} [\ln (mc)]^n + \cdots.
\ee

Recall that we can change our basis of local operators, which is equivalent to
an analytic change of coordinates in the neighborhood of renormalized
trajectories. In other words, this transformation should not spoil the mass
independence of the RG equations

\be
\tilde{\fk}_a  = \fk_a + \sum_{b: x_a>x_b} {n_{a}}^{b}m^{x_a-x_b}\fk_b \equiv
{N_{a}}^{b}(m)\fk_b,
\ee

\nt such that with $\tilde{\g} = N(m)\g N^{-1}(m)$ we can write:

\be \label{vev}
\la \tilde{\fk}_a \ra_m = m^{x_a} \frac{(-1)^n}{n!}
\tilde{\g}_{a}{}^{b_1}\tilde{\g}_{b_1}{}^{
b_2}\cdots\tilde{\g}_{b_{n-1}}{}^{b_n} [\ln (mc)]^n.
\ee

\nt This transformation exists if a reduced matrix, obtained from $\g$ upon
eliminating the entries due to \id, is diagonalizable.

For the RG analysis of more complicated correlation functions we need to know
the matrix $G(r;m)$. In fact we can solve (\ref{green}) very easily to get

\be \label{solgreen}
G(r;m) = m^X r^{-\g} (mr)^{-X},
\ee

\nt where $\g_{a}{}^b$ are non-vanishing only if $x_a-x_b$ is a positive
integer. With the explicit form of $G(r;m)$ at hand, the correlation functions
(\ref{RGns}) are determined up to RG invariant functions. The remainder of this
paper is concerned with perturbative calculation of these RG invariant
functions.
\vskip .75cm
\pagebreak[3]


\section{Perturbation Theory for OPE Coefficients}
\setcounter{equation}{0}
\vskip .25cm
\indent

As we have seen in equation (\ref{Rgcoeffs}), after separating out the
non-analyticities of correlation functions encoded in expectation values of
single operators, we remain with the OPE coefficients which are analytic
functions of parameters. Renormalization group analysis fixes these analytic
functions up to RG invariant combinations (\ref{genH}). Determination of these
functions requires explicit perturbative calculations, based on a concrete
regularization scheme. Note that, since the OPE coefficients are universal,
they do not depend on how we calculate them.

On the other hand, conjugacy in thermodynamics and statistical mechanics is an
exact notion, irrespective of any approximation technique used in practical
calculations. The thermodynamical definition of conjugacy is that the
derivative of a correlation function with respect to a parameter is given by an
insertion of the volume integral of the corresponding conjugate operator. As we
already discussed, well-defined renormalizable field  theories are defined as
the scaling limit of statistical systems. In taking this limit, physical
short-distance singularities appear, which prevent us from naively borrowing
the definition of conjugacy as given in thermodynamics. In order to extend this
definition to field theories, the notion of {\em singular} operator product
expansion is introduced to cure the physical {\em ultraviolet} singularities
mentioned above. Moreover, the {\em full} operator product expansion (both
singular and integrable) need to be introduced to incorporate {\em locality}
into this formalism.

The variational formula for correlation functions gives a realization of the
derivative with respect to a renormalized parameter which satisfies the
requirements imposed by the field-theoretical conjugacy relation\ct{Son4}.
Consistency of this variational formula with the locality of the operator
product expansion amounts to the {\em infrared finiteness} of the variational
formula for the operator coefficients. The latter variational formula has been
introduced and proven to be infrared finite in ref.\ct{Son5}. We provide the
details of the proof of the infrared finiteness to all orders. As a by-product
we arrive at a manifestly finite calculational scheme for the perturbative
calculation of the OPE coefficients. In the following sections we discuss these
issues in detail.

\vskip .25cm
\pagebreak[2]


\subsection{Variational Formula for Correlation Functions}
\vskip .2cm
\nopagebreak
\indent

In order to characterize the short-distance singularities of the renormalized
correlation functions, whose volume integral will be needed shortly, and to
choose the suitable subtractions to make these integrals UV finite, consider
the following singular OPE coefficients

\be \label{OPEs}
\C{O}_i (r) \fk_a (0) = \bar{C}_{ia}^{\scriptscriptstyle{(\! s \!)}} {}^b (r;g)
\fk_b (0)+O(\frac{\displaystyle{1}}{\displaystyle{r^D}}).
\ee

\nt Here, we assume the average over the  orientation of the coordinate vector
$r_ \mu$ has been taken, such that the right hand side includes only the scalar
composite operators. Also note that only composite fields $\fk_b$ whose scale
dimension are at most \,$x_i + x_a - D$ should be kept; we only need to
consider the unintegrable part of the OPE coefficients for calculating the
subtractions.

\indent The divergences arising from the singular OPE coefficients are {\rm
physical} short-distance singularities and they exist even after
renormalization. By using these singular OPE coefficients, the following
variational formula has been introduced\ct{Son4}

\ba \label{VFgn}
\lefteqn{- \pr{i} \la \fk_{a_1}(r_1) \cdots \fk_{a_n}(r_n) \ra_g =\had{\ep}{0}
\lB \Dant{|r' - r_k| \geq \ep}{r'} \la \C{O}_i^c(r') \fk_{a_1}(r_1) \cdots \fk
_{a_n}(r_n) \ra_g }  \nonu \\
& &~~~~~~~~~~~~~~~~~~~~~~~~~~~~~~~~~~~~~~~~~~~~~+ \sum_{k=1}^{n} {A_{ia_k}}^b
(g;\ep) \la \fk_{a_1}(r_1) \cdots \fk_b (r_k) \cdots \fk_{a_n}(r_n) \ra_g
\rB,~~~~~
\ea

\nt where $\pr{i} \equiv \var{g_i}$, the definition of {\em connected}
operators is

\be \label{connected}
\C{O}_i^c (r) \equiv \C{O}_i (r) - \la \C{O}_i \ra_g,
\ee

\nt and the definition of the subtractions, in the matrix notation, is

\be \label{sub}
A_{i}(\ep;g) \equiv - \Dant{1 \geq r \geq \ep}{r}
\bar{C}_i^{\scriptscriptstyle{(s)}} (r;g) + c_{i}(g).
\ee

\nt Here $1$ is the renormalization point, fixed under the renormalization
group.

The arbitrariness involved in the subtraction of the divergent counterterms is
compensated for by the introduction of finite counterterms $c_i (g)$. These
finite counterterms, in addition to the above mentioned {\em physically}
important role, have a natural {\em geometrical} meaning in the theory space.
The correlation functions are tensor fields on the space of renormalized
trajectories, and therefore their partial derivatives are not covariant
quantities. To be able to define a covariant derivative we need to have a
connection on the space of renormalized trajectories in theory space. In other
words, covariance of the VF demands the introduction of finite counterterms
$c_i (g)$ which transform as a connection

\be \label{gatr}
c_i(g) \rightarrow {c^{\prime}_i}(g) = N(g)(\pr{i} + c_i (g))N^{-1}(g),
\ee

\nt under the change of the basis given by equation (\ref{N}). This is how
connection $c_i(g)$ arises naturally in the space of renormalized
trajectories\ct{Son8}.
\vskip .5cm
\pagebreak[2]


\noindent \underline{Consistency Conditions of VF with RG Equations and Maxwell
Relations}
\nopagebreak
\indent

It is crucial to notice that the above VF is assumed to be valid for finite
values of the parameters ${g^i}$, \ie to be valid beyond perturbation theory.
This is a reflection of the non-perturbative nature of the conjugacy relation
even in field theory. Though the definition of the variational formula is
inspired by its statistical mechanical counterpart, the relation between a
lattice model and the corresponding field theory is non-trivial. Consequently,
it is difficult to derive the VF from first principles. Instead, if the
existence of the conjugate fields $\C{O}_i$, satisfying the conjugacy relation
(\ref{con}) and VF (\ref{VFgn}) is assumed, then the consistency of VF can be
checked.  By demanding consistency with the RG equations (\ref{beta}),
(\ref{RG1}), and (\ref{RG1c}) we arrive at the following remarkable
relation\ct{Son4}

\be \label{consist1}
{\rm vol}(S^{D-1})~H_i^{\scriptscriptstyle{(s)}}(g) = \var{g^i} \Psi(g) +
[c_i(g),\Psi(g)] +\bb^j(g) \ob_{ji}(g),
\ee

\nt where

\be \label{Psi}
\Psi(g) \equiv \gb(g)+\bb^i(g)c_i(g),
\ee

\nt is a covariant tensor field,

\be \label{Hs}
H_i^{\scriptscriptstyle{(s)}}(g) \equiv \bar{C}_i^{\scriptscriptstyle{(s)}}
(r=1;g),
\ee

\nt and the curvature $\ob_{ij}$ is defined by

\be \label{cur}
\ob_{ij}(g) \equiv \partial_i c_j (g)- \partial_j c_i (g)+ [c_i (g),c_j (g)].
\ee

The curvature could be determined by demanding the consistency of the VF, when
used recursively to evaluate higher order derivatives. This means that we want
the VF to satisfy the Maxwell relations:

\be \label{max}
\pr{i} \pr{j} = \pr{j} \pr{i}.
\ee

\nt Imposing these conditions on the second order VF for correlation functions
gives an expression for the curvature in terms of a double integral over a {\em
finite} domain\ct{Son4}:

\be \label{curint}
\ob_{ij}(g) \la {\fb} \ra_g = \Dant{1 \geq r}{r} {\rm F.P.} \Dant{1 \geq
r'}{r'} \la \C{O}_j^c (r) \left( \C{O}_i^c (r')
-C_i (r') \right) {\fb}(0)-(i \leftrightarrow j) \ra_g ,
\ee

\nt where F.P. denotes taking the integrable part with respect to $r$. The
commutativity condition for more than two derivatives reduces to the Maxwell
relations (\ref{max}) and does not give any new conditions.
\indent

 Consistency condition (\ref{consist1}) gives a geometrical expression for the
singular part of the OPE coefficients; the vector field $\bb^i(g)$ (beta
functions), the rank-two tensor field $\Psi(g)$ (scale and anomalous
dimensions), and the connection $c_i (g)$ (finite counterterms), determine the
unintegrable part of the short-distance singularities of the theory. These are
analogues of the well-known relation between the renormalization constants, the
beta functions, and the anomalous dimensions in the dimensional regularization
with minimal subtraction scheme\ct{thooft}.
\vskip .25cm
\pagebreak[2]


\subsection{Variational Formula for OPE Coefficients}
\vskip .2cm
\nopagebreak
\indent

Even though the above VF for correlation functions is assumed to be true
non-perturbatively (for arbitrary values of $g$), it is hard to use in
practical calculations. However it is naturally suited for calculating
off-critical correlation functions around a non-trivial fixed point by using
only the information about the correlation functions at the fixed point
($g=0$), where they are easier to calculate because of invariance under scale
and sometimes larger symmetry group transformations. This is nothing other than
perturbation theory around a UV fixed point, and therefore requires the
introduction of a realization of derivatives with respect to the parameters of
the theory, valid particularly at the point $g=0$\footnote{Usually the
correlation functions are known not only for the fixed point but also for some
special path out of the fixed point. In these cases we can use VF on the
trivial path to calculate the correlation functions in its vicinity
perturbatively (\eg In the massive $(\fk^4)_4$ theory, since all correlation
functions are known for the free massive theory, the VF for $\lambda=0$ could
be used to peturbatively calculate the OPE coefficients in the vicinity of the
free massive path \ct{Son5}).}.

The non-analytic dependence of the correlation functions on the parameters
shows up in the VF as IR divergences at $g=0$. But by virtue of (\ref{RGns})
all the non-analytic dependence of correlation functions are encoded in
one-point functions. Therefore we have to look for a VF for OPE coefficients
which, by assumption, are analytic in parameters and perturbatively calculable.
This way of doing perturbative calculations is in the spirit of ref. \ct{Wil1}
and more general than the standard diagrammatic way.

In ref. \ct{Son5}, using the VF of correlation functions, a VF formula for the
OPE coefficients in the product of two operators was given. Along the same line
using the VF (\ref{VFgn}) for the derivatives of correlation functions on both
side of OPE (\ref{OPEn}), we obtain the following VF for general OPE
coefficients\footnote{For the rest of this section all formulas are for a
finite $g$, but for simplicity we suppress some of the $g$-dependences.}

\ba \label{VFcn}
\lefteqn{- \pr{i} {C_{a_1 \cdots a_n }}^b (r_1 , \cdots ,r_n ;g) \la \fk_b
\ra_g = } \nonu \\
& &~~~~~~~~~~~~\had{\ep}{0} \lB \Dant{{\scriptstyle |{r'}-r_i | \geq {\ep}
\atop r' \geq \ep}}{r'} \la \C{O}_i^c (r') \left( \fk_{a_1}(r_1) \cdots
\fk_{a_n}(r_n) - {C_{a_1 \cdots a_n}}^b(r_1, \cdots, r_n ) \fk_b (0) \right)
\ra_g \nonumber \\
& & + \Big(\sum_{k=1}^n {A_{ia_k}}^d ( \ep ) {C_{a_1 \cdots a_{k-1} d a_{k+1}
\cdots  a_n}}^b(r_1, \cdots, r_n)  - {C_{a_1 \cdots a_n }}^d (r_1 , \cdots
,r_n)A_{id}{}^b(\ep) \Big) \la \fk_b \ra_g \rB.~~~~~~~
\ea

To be consistent with the differentiability of OPE coefficients, the above VF
should be free of IR divergences. This consistency is automatic because from
OPE (\ref{OPEn}), for any $ R'> {\rm max} \{ r_1, \cdots r_n \} $, we have

\be \label{IRid1}
\Dant{r' \ge R'}{r'} \la \C{O}_i^c (r') \left( {\fk}_{a_{1}}(r_{1}) \cdots
{\fk}_{a_{n}}(r_{n})-{C_{a_{1} \cdots a_{n}}}^b(r_{1}, \cdots, r_{n})
\fk_{b}(0) \right) \ra_g =0.
\ee

\nt So if we consider $R'$ as the IR cutoff for equation (\ref{VFcn}), the
integral is independent of the cutoff. On the other hand, since the integral is
performed over a finite domain, it is free of infrared divergences as well. As
expected physically, this guarantees that UV physics completely determines the
structure of the OPE coefficients; $R'$ can be chosen as small as, but not
equal to the largest argument of the OPE coefficients.

The VF (\ref{VFcn}) easily generalizes to higher-point functions. We have the
following first order VF for general OPE coefficients when ${\rm max} \{ r_1
\cdots r_n \} < R' < {\rm min} \{ s_1 \cdots s_l \}$

\ba \label{VFcg}
\lefteqn{-\pr{i} {C_{a_{1} \cdots a_{n}}}^b (r_{1}, \cdots , r_{n} ;g) \la
\fk_b (0) \fk_{d_1}(s_1) \cdots \fk_{d_l}(s_l) \ra_g =}  \nonumber \\
& & \!\!\!\!\!\!\!\!\had{\ep}{0} \lB \Dant{\scriptstyle |{r'}-r_i | \geq {\ep}
\atop R' \geq r' \geq \ep}{r'} \la \C{O}_i^c (r') \left( \fk_{a_1}(r_1) \cdots
\fk_{a_n}(r_n) - {C_{a_1 \cdots a_n}}^b(r_{1}, \cdots, r_{n}) \fk_b (0) \right)
\fk_{d_1}(s_1) \cdots \fk_{d_l}(s_l) \ra_g \nonumber \\
& & \!\!\!\!\!\!\!\!+ \Big( \sum_{k=1}^n {A_{ia_k}}^d ( \ep ) {C_{a_1 \cdots
a_{k-1} d a_{k+1} \cdots  a_n}}^b(r_{1}, \cdots, r_{n}) \!-\! {C_{a_1 \cdots
a_n }}^d (r_1 , \cdots ,r_n)A_{id}{}^b(\ep) \Big) \la \fk_b(0) \fk_{d_1}(s_1)
\cdots \fk_{d_l}(s_l) \ra_g \rB,\nonumber\\
\vspace{ -.25cm}
& &
\ea

\nt which is manifestly free of any divergences (both UV and IR).
\vskip .5cm
\pagebreak[2]


\nt \underline{Higher Order VF and Consistency with Analyticity of OPE
Coefficients}
\nopagebreak
\indent

We can use the VF recursively to evaluate higher order derivatives of OPE
coefficients. The consistency of the higher order VF with the commutativity of
the partial derivatives (\ie Maxwell relations) has already been checked and
resulted in an expression for the curvature. This type of consistency is
related to the UV behavior of the variational formula; the curvature is in
terms of double integral over a {\em finite} domain. But we have another
consistency check: the VF should be consistent with the analyticity assumption
of the OPE coefficients. This is related to the IR behavior of VF. In the
following we will show that, like the first order VF, this consistency can be
made automatic for higher orders recursively.
\indent

To avoid too many indices and to see the structure of higher order VF in a more
transparent way, let us consider the OPE coefficients for two fields in the
matrix form. Later we will only use this higher order VF for the Ising model.
Consider the full OPE for two composite fields

\be \label{OPE2}
\fk_a (r) \fb(0) = C_{a}(r;g) \fb (0).
\ee

Consider the following two examples of first order, manifestly finite VFs for
$C_a$:

\ba \label{VFOPE2}
\lefteqn{- \pr{i} C_a (r;g) \la \fb \ra_g =\had{\ep}{0} \lB \Dant{|r'-r| \geq
\ep \atop R' \geq r' \geq \ep}{r'} \la \C{O}_i^c (r') \left( \fk_a (r)-C_a (r)
\right) \fb (0) \ra_g } \nonu \\
& &~~~~~~~~~~~~~~~~~~~~~~~~~~~~~~~~~~~~~~~~~~~~~~~~~~~~~~~~ + \left( [A_i
(\ep),C_a (r)] +{A_{ia}}^d(\ep)C_{d}(r) \right) \la \fb \ra_g \rB,~~~~~~~
\ea
\ba \label{VFOPE22}
\lefteqn{- \pr{i} C_a (r;g) \la \fb (0) \fk_d(s)\ra_g =\had{\ep}{0} \lB
\Dant{|r'-r| \geq \ep \atop R' \geq r' \geq \ep}{r'} \la \C{O}_i^c (r') \left(
\fk_a (r)-C_a (r) \right) \fb (0) \fk_d (s) \ra_{g}}  \nonumber \\
& &~~~~~~~~~~~~~~~~~~~~~~~~~~~~~~~~~~~~~~~~~~~~~~~~~~~~~~~+ \left( [A_i
(\ep),C_{a}(r)] +{A_{ia}}^d(\ep)C_{d}(r) \right) {\la} \fb (0) \fk_d (s) \ra_g
\rB,~~~~~~~
\ea

\nt where $r<R'<s$.

The second order VF could be derived by taking the derivative of VF
(\ref{VFOPE2}) and using the VF for correlation functions (\ref{VFgn}). At the
end we have

\ba \label{VFOPEo2}
\lefteqn{\pr{j} \pr{i} C_a (r;g) \la \fb \ra_g = \had{\ep}{0} \lB \Dant{|r'-r|
\geq \ep \atop R' \geq r' \geq \ep}{r'} \had{\et}{0} \bigg( \Dant{|r''-r| \geq
\et \atop {|r''-r'| \geq \et \atop R'' \geq r'' \geq \et}}{r''} \la \C{O}_j^c
(r'') \C{O}_i^c (r') \Big( \fk_a (r)-C_a (r) \Big) \fb(0) \ra_g} \nonu \\
& & + \left( \pr{j}C_a (r)+[A_j (\et) ,C_a (r)]+{A_{ja}}^b (\et)C_b (r) \right)
\la \C{O}_i^c (r') \fb (0) \ra_g +{A_{ji}}^b (\et) \la \fk_b^c (r') \Big( \fk_a
(r)-C_a (r) \Big) \fb (0) \ra_g \nonu \\
& & +{A_{ja}}^b (\et) \la \C{O}_i^c (r') \Big( \fk_b (r)-C_b (r) \Big) \fb (0)
\ra_g +A_j (\et) \la \C{O}_i^c(r') \Big( \fk_a (r)-C_a (r) \Big) \fb (0) \ra_g
\bigg) \nonu \\
& &  - \pr{j} \left( [A_i(\ep) ,C_a (r)]+{A_{ia}}^b (\ep)C_b (r) \right) \la
\fb \ra_g + \left( \pr{i}C_a (r)+[A_i(\ep) ,C_a (r)]+{A_{ia}}^b (\ep)C_b (r)
\right) \nonu \\
& & ~~~~~~~~~~~~~~~~~~~~~~~~~~~~~~~~~~~~~~~~~~~~~~~~\times \had{\et}{0} \bigg(
\Dant{R'' \geq r'' \geq \et}{r''} \la \C{O}_j^c (r'') \fb (0) \ra +A_j(\et) \la
\fb \ra_g \bigg) \rB,~~~~~~
\ea

\nt where $r<R'<R''$. It is important to note that in the above formula, and
all higher order VFs, whenever two or more connected operators are inserted,
partial connectedness with respect to all of the inserted operators is implied.

The $R'$ independence of VF (\ref{VFOPEo2}) is guaranteed since our starting
point, the first order VF (\ref{VFOPE2}), is already shown to be independent of
$R'$ as a direct consequence of the OPE itself (\ie zero-th order VF). In the
same spirit we can use the first order VF contracted with two point functions
(\ref{VFOPE22}), to find the following identity

\ba
\lefteqn{\had{\ep}{0} \lB \Dant{|r'-r| \geq \ep \atop R' \geq r' \geq \ep}{r'}
\Dant{r''>R''}{r''} \la \C{O}_j^c (r'') \C{O}_i^c (r') \Big( \fk_a (r)-C_a (r)
\Big) \fb(0) \ra_g } \nonu \\
& &~~~~~~~~~~~~~~~~ + \left( \pr{i}C_a (r)+[A_i(\ep) ,C_a (r)]+{A_{ia}}^b
(\ep)C_b (r) \right) \Dant{r'' > R''}{r''} \la \C{O}_j^c (r'') \fb (0) \ra_g
\rB =0. ~~~~~~~~~
\ea

\nt This implies the $R''$ independence of the VF (\ref{VFOPEo2}). So we end up
with a second order VF which is independent of IR cutoffs and its integrals are
over finite ranges, and therefore {\em manifestly} finite.

The above arguments for IR finiteness can be generalized, in a straightforward
manner, to higher order VFs for general OPE coefficients recursively. Generally
for $n^{\underline {th}}$ order VF of OPE coefficients we end up with $n$
nested volume integrals with increasing cutoffs from the outer integrals to the
inner ones. The resulting expression is independent of the cutoffs, the
smallest of which may be chosen as small as, but not equal to the largest
arguments of the OPE coefficient. This proves the differentiability, {\em to
any order}, of all OPE coefficients starting from the VF. In another words,
this shows the consistency of the VF with the analyticity of OPE coefficients
with imposing a mild but calculationally crucial new condition on the
nestedness of the IR cutoffs; this variational formula respects the locality of
the theory.
\vskip .25cm
\pagebreak[2]


\subsection{Systematic Perturbation Theory around a Fixed Point}
\vskip .2cm
\nopagebreak
\indent

Higher order VFs such as (\ref{VFOPEo2}) give us a compact equation for the
derivatives of the OPE coefficients. But, we need more refinements to develop a
systematic method for perturbative calculations\footnote{The notion of
conjugacy in theories with conjugate operator(s) which are singular at $g=0$ is
much more intricate\ct{Son4}.}. First, we must show how to truncate the
infinite sums in (\ref{VFOPEo2}). Second, since calculating the integrals with
finite IR cutoffs is generally difficult, we will remove the cutoffs carefully.
In this section, the above two refinements will be made simultaneously. It then
remains to show that the VFs provide us with sufficient conditions for
calculating the relevant quantities. We will explicitly show this to be true
for the IM in section 5.
\vskip .5cm
\pagebreak[2]


\nt \underline{Truncation of VF for OPE coefficients at the Fixed Point}
\nopagebreak
\indent

On physical grounds, we expect that, in the perturbative calculations around
the point $g=0$, the contribution of operators with high enough scale
dimensions to variational formula for OPE coefficients should die off in the $g
\rightarrow 0$ limit. To show that VFs for OPE coefficients, like equation
(\ref{VFOPE2}), have this property we need to go back to their derivations. To
keep the presentation simple, let us again consider the case of the second
derivative of OPE for two fields. First note that scale covariance
(\ref{scaling}) at the fixed point gives

\ba \label{id}
\la \fk_a \ra =0&~~~~&{\rm if}~~\fk_a \neq \id.
\ea

\nt Here and in what follows we use the convention that fixed point correlation
functions are denoted without any subscripts. Taking derivatives of
(\ref{OPE2}) we have

\be \label{trun}
\pr{j} \pr{i} C_{ab}{}^d (r;g) \la \fk_d \ra_g=\pr{j} \pr{i} \la \fk_a (r)
\fk_b(0) \ra_g - \pr{i} C_{ab}{}^d (r;g) \pr{j} \la \fk_d \ra_g - \pr{j}
C_{ab}{}^d (r;g) \pr{i} \la \fk_d \ra_g - C_{ab}{}^d (r;g) \pr{j} \pr{i} \la
\fk_d \ra_g.
\ee

Because of the basic analyticity assumption for the OPE coefficients the left
hand side of equation (\ref{trun}) is analytic in $g$ except for $\la \fk_d
\ra_g$. So, using (\ref{id}) the right hand side is regular at $g=0$; all
divergences of the terms on the right hand side {\em exactly} cancel one
another. Hence, (\ref{trun}) is well defined even at the fixed point. If we use
the VF for the derivatives in equations (\ref{trun}), we will get the second
order VF (\ref{VFOPEo2}) with the infinite sums over operators with arbitrarily
high scale dimensions. Instead, let us see what we can say about the
derivatives of one point functions before using the VF. Translational
invariance and dimensional analysis considerations imply

\ba \label{dopf}
\pr{i} \la \fk_a \ra =0&~~~~&{\rm if}~~x_a>y_i, \\ \label{dopfb}
\pr{j} \pr{i} \la \fk_a \ra =0&~~~~&{\rm if}~~x_a>y_i +y_j.
\ea

\nt Using the VF for correlation functions, equations (\ref{dopf}) and
(\ref{dopfb}) give us a hierarchy of identities involving single and double
volume integrals over correlation functions. These integrals do not have any IR
cutoffs. Each of these integrals could be split into a main part with an IR
cutoff and a correction part which is over the region beyond the IR cutoff. We
can use these identities to truncate the VF (\ref{VFOPEo2}) such that only a
finite number of terms remain. These identities also provide some restrictions
on the choice of finite counterterms\ct{MZ2}.

If we consider well-behaved field theories in which we only have a finite
number of composite operators with scale dimensions less than any given finite
number, then we are only left with a finite number of terms in the sums and all
other terms could be written in terms of these corrections. Because of the
dimensional restrictions and the connectedness of the insertions, the
corrections vanish at $g=0$ when we take the nested infinite limit of the
cutoffs (from the inner to the outer ones). After all these steps we end up
with the following VF at the fixed point

\ba \label{tVFOPEo2}
\lefteqn{\pr{j} \pr{i} C_{ab}{}^{\id} (r) = \had{R'}{\infty} \had{R''}{\infty}
\Bigg{\{}  r^{(y_i+y_j-x_a-x_b)} \C{M}_{ij,ab}\Big(
\frac{R'}{r},\frac{R''}{r}\Big)  -r^{(y_i+y_j-x_{d_{ij}})} C_{ab}{}^{d_{ij}}
(r) \C{S}_{ij,d_{ij}}^{IR} \Big( \frac{R'}{r},\frac{R''}{r}\Big) } \nonu \\
& &~~~~~~~~~~~~~~~~~~~~~~~~~~~~~~~~~~~~~~~~~~~+\bigg( r^{(y_i-x_{d_i})}
\pr{j}C_{ab}{}^{d_i} (r) \C{S}_{i,d_i}^{IR}\Big( \frac{R'}{r}\Big) -\Big(
(i,R') \leftrightarrow (j,R'')\Big) \bigg) \Bigg{\}},~~~~~~~
\ea

\nt where the sums on the repeated indices are restricted by

\ba \label{dtrun}
&x_{d_i} \leq y_i,& \\
&x_{d_{ij}} \leq y_i +y_j,&
\ea

\nt and the main term and the two types of IR subtraction integrals are defined
as follows:

\ba \label{Mn}
\lefteqn{\C{M}_{ij,ab}(U',U'')=} \nonu \\
& &r^{(x_i+x_j+x_a+x_b)}\had{\ze}{0} \lB \Dant{|u'-1| \geq \ze \atop U' \geq u'
\geq \ze}{u'} \had{\xi}{0} \bigg( \Dant{|u''-1| \geq \xi \atop {|u''-u'| \geq
\xi \atop U'' \geq U'' \geq \xi}}{u''} \la \C{O}_j^c (ru'') \C{O}_i^c (ru')
\fk_a (r) \fk_b (0) \ra
 \nonu \\
& &+ A_{ji}{}^e (r\xi) \la \fk_e^c (ru') \fk_a (r)\fk_b (0) \ra+ A_{ja}{}^e
(r\xi) \la \C{O}_i^c (ru') \fk_e (r) \fk_b (0) \ra +A_{jb}{}^e (r\xi) \la
\C{O}_i^c (ru') \fk_a (r) \fk_e (0) \ra \bigg) ~\nonu \\
& &- \pr{j} \Big( A_{ia}{}^e (r\ze) C_{eb}{}^{\id} (r)+A_{ib}{}^e
(r\ze)C_{ae}{}^{\id} (r) \Big)+ \Big(A_{ia}{}^e (r\ze)C_{eb}{}^{d_j}
(r)+A_{ib}{}^e (r\ze)C_{ae}{}^{d_j} (r) \Big) \nonu \\
& & ~~~~~~~~~~~~~~~~~~~~~~~~~~~~~~~~~~~~~~~\times \had{\xi}{0} \bigg(
\Dant{{\scriptscriptstyle U'' \geq u'' \geq \xi}}{u''} \la \C{O}_j^c (ru'')
\fk_{d_j} (0) \ra +A_{jd_j}{}^{\id}(r\xi) \bigg) \rB \Bigg{\}},
\ea
\ba \label{Sid}
\C{S}_{i,d}^{IR}(U') =   r^{(x_i+x_d)} \had{\ze}{0} \lB \Dant{|u'-1| \geq \ze
\atop U' \geq u' \geq \ze}{u'} \la \C{O}_i^c (ru') \fk_{d} (0) \ra
+A_{id}{}^{\id}(r\ze) \rB,
\ea
\ba \label{Sijd}
\lefteqn{\C{S}_{ij,d}^{IR} (U',U'') = r^{(x_i+x_j+x_a+x_b)} \had{\ze}{0} \lB
\Dant{|u'-1| \geq \ze \atop U' \geq u' \geq \ze}{u'} \had{\xi}{0} \bigg(
\Dant{|u''-1| \geq \xi \atop {|u''-u'| \geq \xi \atop U'' \geq U'' \geq
\xi}}{u''} \la \C{O}_j^c (ru'') \C{O}_i^c (ru') \fk_a (r) \fk_d (0) \ra} \nonu
\\
& &+A_{ji}{}^e (r\ze) C_{ab}{}^{d} (r)\la \fk_e^c (ru') \fk_{d} (0) \ra
+C_{ab}{}^{d} (r) A_{jd}{}^e (r\xi) \la \C{O}_i^c (ru') \fk_e (0) \ra \bigg)
\nonu \\
& &-C_{ab}{}^{d} (r) \pr{j} A_{id}{}^{\id}(r\ze)-C_{ab}{}^{d}
(r)A_{id}{}^{e}(r\ze) \had{\xi}{0} \bigg( \Dant{{\scriptscriptstyle U'' \geq
u'' \geq \xi}}{u''} \la \C{O}_j^c (ru'') \fk_{e} (0) \ra +A_{je}{}^{\id}(\ze)
\bigg) \rB \Bigg{\}},~~~~~~~~
\ea

\nt where the range for the sums on the upper indices of the subtractions ($e$
in the above equations) is restricted by the dimensional constraint in the
singular OPEs (\ref{OPEs}). Therefore, (\ref{tVFOPEo2}) is a {\em finite} (both
UV and IR) and {\em truncated} variational formula at {\em the fixed point.\/}
After taking the limits as specified above, we do not have any cutoffs left,
neither UV nor IR. The generalization of the above construction to higher
orders is straightforward. It is, however, important to notice that the order
of the limits is crucial; they do not commute. In addition, it is important
that we only need to calculate the integrals for infinitely large IR and small
UV cutoffs. The case of the Ising model is an instructive example of these
points.
\vskip .5cm
\pagebreak[2]


\nt \underline{Calculational Scheme}
\nopagebreak
\indent

The generality of the finiteness proof, presented above, assures the validity
of the following shortcut for writing a truncated variational formula at the
fixed point. We truncate the formal expression obtained by repeated
differentiation of the OPE and its evaluation at the fixed point. Without loss
of generality, consider the case of second order perturbation. From equations
(\ref{trun}), (\ref{dopf}), and (\ref{dopfb}) we find

\be \label{trun0}
\pr{j} \pr{i} C_{ab}{}^{\id} (r)=\pr{j} \pr{i} \la \fk_a (r) \fk_b(0) \ra -
\pr{j} C_{ab}{}^{d_i} (r) \pr{i} \la \fk_{d_i} \ra - \pr{i} C_{ab}{}^{d_j} (r)
\pr{j} \la \fk_{d_j} \ra-C_{ab}{}^{d_{ij}} (r) \pr{j} \pr{i} \la \fk_{d_{ij}}
\ra.
\ee

\nt Now, we replace the derivatives of the correlation functions with their
respective variational formulas in the following special way. Every $\pr{i}$
represents an integration over the position of the inserted operator $\C{O}_i$.
The range of integration should be bounded by the same IR cutoff, respecting
the nestedness of the cutoffs in the main term. So we arrive at the second
order finite and truncated VF (\ref{tVFOPEo2}), directly.

Then, after having the VFs up to the desired order, we must go through the
following steps. First, we should calculate the UV subtractions (\ref{sub}) in
terms of beta functions, anomalous dimensions, and finite counterterms using
the consistency relation (\ref{consist1}) and the solution of the RG equation
(\ref{Rgcoeffs}) for the singular OPE coefficients\ct{Son5}. Second, we must
calculate all the necessary critical correlation functions. This can be hard
for non-trivial fixed points, and is usually only possible for fixed point
theories invariant under a large symmetry group. Third, we can calculate the
volume integrals over these correlation functions with infinitely large IR and
small UV cutoffs. This is usually the hardest step.  We explicitly implement
these steps for the spin-spin correlation function in the Ising model to all
orders of perturbation theory. Finally, substituting the above relations into
the derived VFs, we end up with a set of algebraic equations in terms of the
scaling OPE coefficients, their corrections, anomalous dimensions of the
composite operators, and the finite counterterms. After choosing a convention
for off-critical composite operators by fixing the finite counterterms, up to
the restrictions given by the explicit expression for the curvature
(\ref{curint}), we expect to have sufficient information to solve these
equations. A perturbative expansion of the short-distance expansion of
correlation functions follows. This step has been carried out up to third order
in our previous work\ct{MZ1}.

Let us close this section by addressing a subtle point in using the VF at a
fixed point. There are cases where one or more derivatives of some one point
functions are non-analytic at the fixed point, but the naive $g \rightarrow 0$
limit of the terms corresponding to these derivatives inside the VF give zero
due to the critical OPE algebra. The reason for this {\em apparent\/}
inconsistency is that, in order to capture the correct non-analyticity
structure of the VF for the critical correlation functions, we must be careful
about the relation between the two limits, $g \rightarrow 0$ and $R \rightarrow
\infty$, where $R$ is a typical IR cutoff. However, because of our proof for
the finiteness of the VF for OPE coefficients, we can take the $g \rightarrow
0$ limit of the correlation functions before evaluating the integrals; the
nested structure of the IR limits eliminates the kind of divergences which
ordinarily appear as singular expressions. In section 5, we will verify this
point in the Ising model.
\vskip .75cm
\pagebreak[3]


\section{UV Structure of VF for the Ising Model}
\setcounter{equation}{0}
\vskip .25cm
\nopagebreak
\indent

In the rest of the paper we concentrate on using VF for the calculation of
various off-critical quantities for the two dimensional Ising model (IM). Since
there are no dimensionless parameters in IM (\ie the theory is
super-renormalizable) the structure of UV divergences is relatively
simple\footnote{For a non-trivial example, as far as UV divergences are
concerned, consult ref. \ct{Son5} in which the VF was used to study the
perturbatively renormalizable $(\fk^4)_4$ theory.}. On the other hand, for spin
fields the critical IM is not trivial and this makes the calculations of the
finite parts and the structure of IR divergences of physical quantities very
interesting and non-trivial. In this respect IM shares some of the
complications of higher order calculations in other off-critical theories in
higher dimensions and especially those in two dimensions near unitary minimal
models.
\vskip .5cm
\pagebreak[2]


\nt \underline{Singular OPE Subtractions}
\nopagebreak
\indent

For determining the necessary subtractions in the VF (with respect to $m$), we
need the following singular OPE coefficients

\be \label{IOPEs}
\C{O}_m (r) \fk_a (0) = \bar{C}_{ma}^{\scriptscriptstyle{(s)}} {}^b (r;m) \fk_b
(0)+O(\frac{\displaystyle{1}}{\displaystyle{r^2}}).
\ee

\nt From the general solution for OPE coefficients (\ref{Rgcoeffs}), the
singular part of $C_m$ is given by

\be \label{sIOPEs}
C_m^{\scriptscriptstyle{(s)}} (r;m)= \frac{1}{r}G(r;m)
H_m^{\scriptscriptstyle{(s)}}(mr) G^{-1}(r;m),
\ee

\nt where the analyticity of OPE coefficients implies that $H_m^{(s)}(m)$ is a
power series in $m$. This together with (\ref{solgreen}) and the fact that the
singular part of the OPE coefficients must be at least as singular as $1 \over
r^2$ lead to $H_m^{(s)}(m)$\ct{Son4} as a finite polynomial in $m$:

\be
H_{ma}^{\scriptscriptstyle{(s)}}{}^b(m) = \sum_{k=0}^{x_a -x_b -1}
\frac{m^k}{k!} ~\pr{m} H_{ma}^{\scriptscriptstyle{(s)}}{}^b(0).
\ee

On the other hand, from the consistency relation (\ref{consist1}) we have

\be \label{consI}
2 \pi H_m^{\scriptscriptstyle{(s)}}(m)= \pr{m} \gb (m)+c_m(m)+m \pr{m}c_m
(m)+[c_m (m), \gb (m)],
\ee

\nt where we have used $\ob_{mg_{\scriptscriptstyle{\id}}}(m) =0$. Inserting
(\ref{consI}) in (\ref{sIOPEs}), we can write the singular OPE coefficients in
terms of the finite counterterms and anomalous dimensions. To calculate the
subtractions we need to calculate the volume integral of the singular OPE
coefficients. Note that the matrix $G(r;m)$ satisfies

\be
r\frac{\partial}{\partial r}G(r;m)=-G(r;m) \gb (mr).
\ee

\nt We can therefore write the singular OPE coefficients as total derivatives

\be
C_m^{\scriptscriptstyle{(s)}} (r;m)= \frac{1}{2 \pi r} \frac{\partial}{\partial
r} \bigg[ G(r;m) \Big( \pr{m}+rc_m (mr) \Big) G^{-1}(r;m)+ \pi \bb_{\id}mr^2
(\ln r- \frac{1}{2}) \bigg].
\ee

\nt This means that we can calculate the integral in (\ref{sub}), and therefore
the subtractions can be given in the following closed form

\be \label{subI}
A_m(\ep;m)=G(\ep;m) \left( \pr{m}+ \ep~c_m (m \ep)\right)G^{-1}(\ep;m).
\ee

\nt Hence, we have an explicit expression for the subtractions in terms of the
matirx $G(r;m)$ and the finite counterterms $c_m (m)$ only.  Recall that the
anomalous dimension matrix $\gb(m)$ (\ref{solgreen}) fixes $G(r;m)$ completely.
Therefore, for the case of Ising model, we have completed the first step in the
calculational scheme which was outlined at the end of section 3.
\vskip .5cm
\pagebreak[2]


\nt \underline{Calculation of the Anomalous Dimension of the Conjugate
Operator}
\nopagebreak
\indent

Let us consider the VF for the composite operator $\Om$. From (\ref{VFgn}) we
have

\be \label{VFOm}
- \pr{m} \la \C{O}_m \ra_m = \had{\ep}{0} \lB \ant{r' \geq \ep}{r'} \la \C{O}_m
(r') \C{O}_m (0) \ra_m^c +{A_{mm}}^{\id}(\ep;m) \rB,
\ee

\nt where from (\ref{subI}) and (\ref{solgreen}) we have the following
expression for the subtraction

\be \label{Amm1b}
A_{mm}{}^{\id}(\ep ;m)=-\bb_{\id} \ln (\ep)+c_{mm}{}^{\id}(m).
\ee

Because of the relationship between the scaling limit of the IM and the theory
of a free massive Majorana fermion, we have the advantage of having a
nonperturbative answer for correlation functions involving only the energy
density operator. The off-critical energy density operator is the most general
scalar operator of scale dimension one, and in the fermionic representation it
is a linear combination of $i\sib \si$ and $m \id$

\be \label{last}
\Om = \frac{1}{2\pi} (i\sib \si) +am \id.
\ee

\nt This fixes $\C{O}_m$ to be $\C{E} = \frac{1}{2\pi} (i\sib\si)$ at the fixed
point. Now from the theory of a free massive Majorana fermion in two dimensions
we have\ct{Itz}

\be \label{FMF}
\la \Om (r) \Om (0) \ra_m^c =
 \la i\sib\si(r) \cdot i\sib\si(0) \ra_m^c
= m^2 \left( {\sl{K}}_1^2 (mr)- {\sl{K}}_0^2 (mr) \right),
\ee

\nt independent of $a$. The integral in the VF (\ref{FMF}) can be done
analytically and by comparing the result with the derivative of (\ref{Om}) with
respect to $m$, we find the anomalous dimension of the conjugate operator to
be\footnote{To find $\bb_{\id}$ we do not need to calculate the integral in VF.
It is enough to compare (\ref{Amm1b}) with the corresponding subtraction coming
from the following singular OPE coefficients\ct{MZ1} \[
\bar{C}^{\scriptscriptstyle{(s)}}_{mm}{}^ \id (r;m) = \frac{1}{4 \pi^2
r^2}.\vspace{-.5cm}\]}

\be \label{bb1}
\bb_{\id}=-\frac{1}{2 \pi},
\ee

\nt and the integration constant appearing in (\ref{Om}), (\ref{vev}), and the
final expression for the free energy density (\ref{Fsol}) to be related to the
finite counterterm as follows

\be \label{ccm}
c = \frac{1}{2}~{\rm exp}[\g - 2 \pi {c_{mm}}^{\id}],
\ee

\nt where $\g$ is the Euler constant. Therefore, upon fixing the finite
counterterm $c_{mm}{}^{\id}$, the integration constant is also fixed.
\vskip .5cm
\pagebreak[2]


\nt \underline{Choice of the Parameters}
\nopagebreak
\indent

The precise definition of the conjugate operator $\Om$ depends on the choice of
the parameters $g_{\id}$ and $m$ which are by no means unique. We can
reparametrize the space of renormalized trajectories by some general mixing of
the old parameters up to two general restrictions \ct{Son3}: The first is that
the scale dimensions of the new parameters should be the same as the old ones,
and be additively conserved under the RG flow. The second is that, in keeping
with the locality of the theory, the reparametrization has to be analytic (\ie
it is given as a power series). So in our case we have only one mixing as
follows

\ba \label{repg}
&\tld{g}_{\id}=g_{\id}+ \frac{\displaystyle{\alpha}}{\displaystyle{2}} m^2,& \\
&\tld{m}=m,&
\ea

\nt where we have kept the normalizations fixed. Under this reparametrization,
the conjugate operator and the only relevant finite counterterm change simply
as

\ba \label{repOm}
&\tld{\C{O}}_m = \Om - \alpha m \id ,& \\
&\tld{c}_{mm}{}^{\id}= {c_{mm}}^{\id}+ \alpha.&
\ea

\nt Hence, the mixing of $\C{O}_m$ with $\id$, or equivalently the change of
parameter $g_{\id}$, can be controlled by fixing the choice of
${c_{mm}}^{\id}$. Note that, as already discussed in section 2, the correlation
functions do not depend on $g_{\id}$. This and the fact that any change in
${c_{mm}}^{\id}$ can be absorbed in a change of $g_{\id}$, amount to the
${c_{mm}}^{\id}$ independence of the correlation functions.

If we consider off-critical correlation functions which involve only operators
corresponding to spin and energy density operators, then for the possible
subtractions needed in the VF, $\fk_a$ in  (\ref{IOPEs}) should be only the
spin operator $\s$ or the conjugate operator $\Om$ (which respectively
correspond to $\s$ and $\C{E}$ at the fixed point). In this case, because of
the dimensional restriction, the only relevant subtraction is the following
which we have already seen in the VF for $\Om$:

\be
{A_{mm}}^ \id (\ep;m) = \frac{1}{2 \pi} \ln (\ep)+{c_{mm}}^\id .
\ee

\nt This is the main reason that the necessary UV subtractions in VFs for IM
are relatively simple, especially in higher orders of perturbation. Generally,
this is a feature of super-renormalizable theories. Consider the case of the
$n^{\underline{th}}$ order VF. This means that we have to consider the
insertion of volume integral of $\C{O}_m^c$s, and the UV divergent subtractions
needed when the inserted operators approach one another or other operators in
the correlation function. But, the UV subtractions corresponding to  two
connected energy density operators are zero. This can be seen in the VF because
of the obvious fact that: $\id^c=0$. In short; most, but not all, of the UV
divergences in higher order perturbative calculations of IM are taken care of
by the connectedness of the insertions. The rest of the subtractions require
the calculation of the other elements of the anomalous dimension matrix
$\Gamma(g)$. $\bb_\id$ is the only element needed for the calculation of the
short-distance expansion of the spin-spin correlation function to third
order\ct{MZ1}.
\vskip .75cm
\pagebreak[3]


\section{Perturbation Theory for the Ising Model}
\setcounter{equation}{0}
\vskip .25cm
\nopagebreak
\indent

In this section we implement the calculational scheme, outlined in section 3,
for perturbative analysis of the spin-spin correlation function in the IM in
the vicinity of the critical point. After studying the structure of the
$n^{\underline{th}}$ VF, we carry out the first two steps of our scheme for the
$n^{\underline{th}}$ order calculation of corrections to scaling. First, we
study the structure of the necessary critical correlations. Then, we present a
systematic method to calculate the necessary integrals over the critical
correlation functions. The final step, which is the fixing of the off-critical
operators and solving the resultant algebraic equations, is done up to third
order. This section is a systematic extension of the explicit third order
calculation done in reference\ct{MZ1}, the details of which are presented in
appendix B. We leave a systematic study of the final step for general
$n^{\underline{th}}$ order corrections to scaling, for a future work\ct{MZ2}.
\vskip .25cm
\pagebreak[2]


\subsection{$n^{\underline{th}}\;$ Order Correction to Scaling}
\vskip .2cm
\nopagebreak
\indent

Consider the short-distance expansion of the spin-spin correlation function.
Due to the OPE algebra at the critical point, using the notation of
ref.\ct{BPZ} we have

\be \label{Cssk}
\la \s (r) \s (0) \ra_m =C_{\s \s}{}^{1 ;\{ b\} }(r;m) \la \id^{\{ b\} } \ra_m
+ C_{\s \s}{}^{\e ;\{ b\} } (r;m) \la \e^{\{ b\} } \ra_m,
\ee

\nt where $\id^{\{ b\} }$ and $\e^{\{ b\} }$ are the off-critical deformations
of the scalar operators in the families $[\id]$ and $[\e]$ (see appendix A). In
the above summations, it is enough to consider only those operators which are
not total derivatives, since the total derivatives have zero expectation value.

The OPE algebra of the IM at the fixed point has a $\IZ_2$ symmetry, known as
Kramers-Wannier duality, which implies that the correlation functions involving
operators in $[\id]$ and an odd number of operators in $[\e]$ are zero. As we
have already argued at the end of section 3, this eliminates the contribution
of the corresponding derivatives of the one point functions in the VF for the
OPE coefficients. Hence, as a natural extension of the results of section 3,
using

\be
\pr{m}^n \la \fk_k \ra =0~~~~~~~~~~~~~~~~~~~~ {\rm for}~~ n \le x_k,
\ee

\nt and keeping the subtlety in cancellation of IR divergences in mind, we
obtain

\ba \label{Iturn0}
\lefteqn{\pr{m}^n C_{\s \s}{}^{\id} (r)= \pr{m}^n \la \s (r) \s (0) \ra} \nonu
\\
& &- \sum_{k=1 \atop k~{\rm  even}}^{n} \left( {n \atop k} \right) \sum_{b
\atop x_b \leq k} \pr{m}^{(n-k)} C_{\s \s}{}^{1 ;\{ b\} }(r) \pr{m}^{k} \la
\id^{\{ b\} } \ra
-\sum_{k=1 \atop k~{\rm  odd}}^{n} \left( {n \atop k} \right) \sum_{b \atop x_b
\leq k} \pr{m}^{(n-k)} C_{\s \s}{}^{\e ;\{ b\} }(r) \pr{m}^{k} \la \e^{\{ b\} }
\ra.
\ea

\nt Note that there are a finite number of operators with $x_b \leq n$, and
therefore there are a finite number of IR subtractions.

Using the short cut developed in section 3 we can write following finite and
truncated $n^{\underline{th}}$ order VF

\ba \label{VFn}
\pr{m}^n C_{\s\s}{}^{\id}(r)&\!\!\!\!=&\!\!\!\! \had{R_1}{\infty} \cdots
\had{R_n}{\infty} \bigg{\{ } (-1)^n r^{n-\frac{1}{4}} \C{M}_n\Big(
\frac{R_1}{r} \cdots \frac{R_n}{r}\Big) \nonu \\
&\!\!\!\!-&\!\!\!\! \sum_{k=1 \atop k~{\rm  even}}^{n} \sum_{b \atop x_b \leq
k} r^{(k-x_{1 ;\{ b\} })} ~\pr{m}^{(n-k)} C_{\s \s}{}^{1 ;\{ b\} }(r)
\sum_{{\rm perm}~\{ p_1,\cdots ,p_k \} \atop {\rm from}~\{ 1,\cdots ,n\} }
\C{S}_{k,1;\{ b\}}^{IR}\Big( \frac{R_{p_1}}{r} \cdots \frac{R_{p_k}}{r}\Big)
\nonu \\
&\!\!\!\!+&\!\!\!\! \sum_{k=1 \atop k~{\rm  odd}}^{n} \sum_{b \atop x_b \leq k}
r^{(k-x_{\e ;\{b \} })} ~\pr{m}^{(n-k)} C_{\s \s}{}^{\e ;\{ b\} }(r) \sum_{{\rm
perm}~\{ p_1,\cdots ,p_k \} \atop {\rm from}~\{ 1,\cdots ,n\} } \C{S}_{k,\e
;{\{ b\} }}^{IR}\Big( \frac{R_{p_1}}{r} \cdots \frac{R_{p_k}}{r}\Big) \bigg{\}
},~~~~~
\ea

\nt where the main term is

\be \label{main}
\C{M}_n(U_1 \cdots U_n) = \ant{\scr{U_1\geq u_1}}{u_1} \cdots \ant{\scr{U_n\geq
u_n}}{u_n} r^{n+\frac{1}{4}}~\la \e(ru_n) \cdots \e(ru_1) \s(r) \s(0) \ra^c,
\ee

\nt and the IR subtraction integrals are

\ba \label{IRS}
\lefteqn{\C{S}_{2k,1;\{ b\} }^{IR}(U_1 \cdots U_{2k}) = } \nonu \\
& & \had{\xi_1}{0} \cdots \had{\xi_k}{0} \Big[ \ant{\scr{U_1\geq u_1 \geq
\xi_1}}{u_1} \cdots \ant{\scr{U_{2k}\geq u_{2k}\geq \xi_{2k}}}{u_{2k}}
r^{(x_{1;\{ b\} } +2k)}\la \e^c(ru_{2k}) \cdots \e^c(ru_1) \id^{\{ b\} }(0) \ra
\nonu \\
& &~~~~~~~~~~~~~~~~~~~~~~~~~~~~~~~~~~~~~~~~~~~~~~~~~~~~~~~~~~~~- \C{S}_{2k,1;\{
b\} }^{UV}(\xi_1, \cdots \xi_{2k},U_1,\cdots ,U_{2k}) \Big],~~~~~~~
\ea

\nt and similar expressions for the IR subtraction integrals $\C{S}_{2k+1,\e;\{
b\} }^{IR}$. The  UV subtractions $\C{S}_{2k,1;\{ b\} }^{UV}$ and
$\C{S}_{2k+1,\e;\{ b\} }^{UV}$, similar to equation (\ref{Sijd}), can be found
in terms of $A_m(\ep)$, its derivatives, and fewer volume integrals over
correlation functions.  The explicit forms of the VFs, up to third order, are
given in appendix B.

The VF (\ref{VFn}) shows the nature of the {\em communication} between UV and
IR divergences explicitly. The precise statement is that the perturbation for
spin-spin correlation function does not have any explicit UV divergences, it
involves IR divergences in $\C{M}_n$ which show up in the end as non-analytic
dependence of the correlation function on the mass. To reveal this structure we
need to handle the IR divergences by using the IR subtraction terms, involving
$\C{S}_{2k,1;\{ b\} }^{IR}$ and $\C{S}_{2k+1,\e;\{ b\} }^{IR}$, which are
naturally given by OPE. But $\C{S}^{IR}$s have both IR divergences (which
cancel the IR divergences of the main term) and UV divergences (cancelled by
the necessary UV subtractions, $\C{S}^{UV}$s).
\vskip .25cm
\pagebreak[2]


\subsection{Calculational Machinery}
\vskip .2cm
\nopagebreak


\nt \underline{Critical Correlation Functions}
\indent

The free massive Majorana fermion representation for the critical Ising
model\ct{BPZ} makes the calculation of the energy-density correlation functions
trivial. Using the relation (\ref{last}) at $m=0$, calculation of correlation
functions involving only the energy-density operators reduces to a product of
Pfaffians\ct{Itz}. Then correlation functions of operators in families $[\id]$
and $[\e]$ could be calculated by using the Virasoro algebra. This takes care
of all the correlation functions we need for IR subtractions. Note, as an
important feature of these correlation functions which makes it possible to
calculate their volume integrals, that they do not involve square roots.

The expression of $\s(r)$ in terms of the fermion field, on the other hand, is
non-local. This makes the calculation of the correlation functions involving
the spin operator non-trivial\footnote{It is mainly due to this that even
though the critical IM has a representation in terms of a free massless
Majonara fermion, it is still a {\em non-trivial\/} fixed point.}. In two
dimensions, due to conformal invariance, all correlation functions are
calculable in principle\ct{BPZ}. Moreover, in the case of the Ising model there
are additional structures which help in principle to write systematically {\em
explicit} recursive answers for the correlation functions of the type needed in
the main term. Duplicating the IM gives a $c=1$ theory which is amenable to
bosonization, and hence the squares of the correlation functions could be
calculated. This relationship was recognized much earlier as the equivalence of
the Ashkin-Teller model at the end of the self dual line to two non-interacting
Ising models\ct{Kad}. Using the above relationships Di Francesco, Saleur, and
Zuber found recursive relations for the correlation functions involving energy
and spin operators\ct{DSZ}. For correlation functions appearing in the main
term (\ref{main}), in terms of a free bosonic field $\fk$ we have

\ba
\lefteqn{\la \e(r_n) \cdots \e(r_1) \s(r) \s(0) \ra \la \s(r) \s(0) \ra =
\frac{1}{\pi^n} \la \cos 2 \fk(r_n) \cdots \cos 2 \fk(r_1) \cos \fk(r) \cos
\fk(0) \ra} \nonu \\
& & ~~~~~~~~~~~~~~~~~~~~~~~~-\frac{1}{2} \sum_{k=1}^{n-1} \sum_{{\rm
permutation~of} \atop{\{p_1, \cdots ,p_n \} \atop {\rm from}~ \{1, \cdots ,n\}
}} \la \e(r_{p_n}) \cdots \e(r_{p_{k+1}}) \s(r) \s(0) \ra \la \e(r_{p_k})
\cdots \e(r_{p_1}) \s(r) \s(0) \ra.~~~~~~~~
\ea

\nt See appendix A for the details and the calculation of correlation functions
needed up to third order.

{}From the above recursive relation and the neutrality condition (\ref{neut})
it can be shown inductively that the connected correlation functions have the
following structure

\be \label{struct}
\la \e(ru_n) \cdots \e(ru_1) \s(r) \s(0) \ra^c = r^{n- \frac{1}{4}}
\prod_{k=1}^n \frac{1}{2\pi |u_k||u_{k}-1|} ~ \prod_{p,q=1 \atop p<q}^n
\frac{1}{|u_p-u_q|^2} ~ F_n(u_1, \cdots,u_n),
\ee

\nt where $F_n(u_1, \cdots,u_n)$ is a real analytic function without any square
roots like $|u_p-u_q|$. Moreover, $F_n$ has the following short-distance
structure

\be
F_n(u_1, \cdots,u_n) \stackrel{u_p \rightarrow u_q}{\longrightarrow}
O(u_p-u_q).
\ee

\nt Hence, the connected correlation function does not have any unintegrable
singularities. Therefore we have the following structure for $\C{M}_n$

\be \label{int}
\C{M}_n(U_1,\cdots ,U_n) = \int_{\scr{U_1 \geq u_1}} \frac{d^2u_1}{2\pi
|u_1||u_1-1|} \cdots \int_{\scr{U_n \geq u_n}} \frac{d^2u_n}{2\pi |u_n||u_n-1|}
{}~\prod_{p,q=1 \atop p<q}^n \frac{1}{|u_p-u_q|^2} ~F_n(u_1,\cdots ,u_n).
\ee

The need to take the IR limits to infinity in (\ref{VFn}), gives us a very
important calculational advantage: usually it is much easier to calculate
integrals for infinitely large cutoffs than for finite ones. This means we are
looking for calculation of $\C{M}_n$ and $\C{S}^{IR}$'s for

\be \label{nejat}
1 \ll U_1 \ll \cdots \ll U_n.
\ee

\nt Even after the above simplification, the calculation of $\C{M}_n$ involves
elliptic functions, and beyond first order is extremely hard to do directly. In
 the following, we develop a systematic way to handle these integrations for
any order.
\vskip .5cm
\pagebreak[2]


\nt \underline{Symmetrization Transformation}
\indent

For each volume integration in $\C{M}_n$ one copy of the integrand of

\be \label{m1}
\C{M}_1(U_1) = \int_{\scr{U_1 \geq u_1}} \frac{d^2u_1}{4 \pi |u_1||u_1-1|}
\ee

\nt appears. According to the above analysis this is the only source of square
roots. Therefore, we can pinpoint the root of the calculational difficulty in
the very first order integral $\C{M}_1$. The key point in overcoming this
difficulty is to use the {\em symmetry\/} properties of the integral. This can
be seen by comparing the integral in (\ref{main}) with the ones in the IR
subtraction (\ref{IRS}); in the IR subtraction terms we have both IR and UV
divergences. More precisely, the structure of the IR subtractions is invariant
under the simultaneous inversion of all variables of the integrals. In the case
of the main term only the $\C{M}_1$-like factors do not respect this symmetry.
The absence of UV divergences in the main term (\ref{main}) is directly related
to separation of the singular points in the $\C{M}_1$-like part of integrand,
one at zero and the other at $1$. Therefore it is enough to look for a
transformation which symmetrizes factors of this type.

This desired transformation changes the complex variable $r_i$ with some linear
function of $x_i+ 1/{x_i}$, where $x_i$ is the new complex variable. Consider
the conformal transformation from $x_i$ to $r_i$

\be \label{symm}
r_i = \frac{r'_i+r''_i}{2}+ \frac{r'_i-r''_i}{4} \Big( x_i+ \frac{1}{x_i}
\Big).
\ee

\vspace{1.25in}
\begin{figure}[h]
	\epsfysize=2.4in
	\hspace{2.75cm}\epsffile[65 250 400 500]{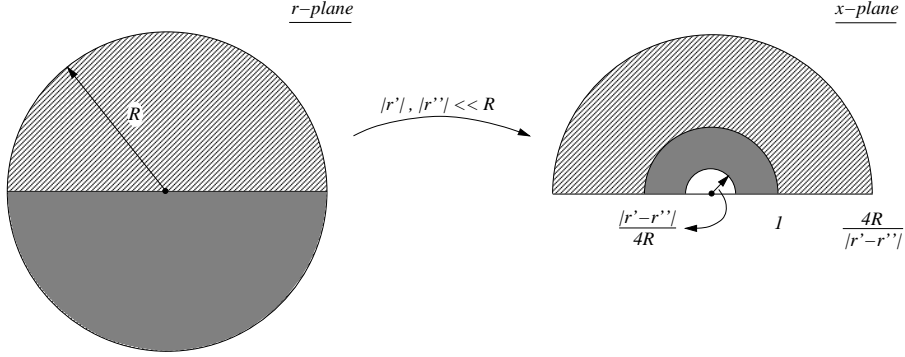}
       \vspace{-1.65in}
       \caption{Symmetrization Transformation}
       \label{fig1}
\end{figure}
\pagebreak[3]

\nt which is called the symmetrization transformation\ct{MZ1}. It merges the
two integrable singularities in $r'_i$ and $r''_i$ to one unintegrable
singularity at zero and enforces the symmetry between UV and IR. We only need
to calculate the integrals for infinitely large cutoffs as it is given in
(\ref{nejat}). This is crucial in having a simple transformation for the IR
cutoffs under the symmetrization (See figure 1). Moreover, using the inversion
symmetry which has been produced by symmetrization, we can restrict the domain
of integration to the region with radius larger than $1$. So we have the
following transformation for a typical integral under symmetrization

\be
\int_{\scr{R_i \geq r_i}} \frac{d^2r_i}{|r_i-r'_i||r_i-r''_i|} =
\int_{\scr{\frac{4R_i}{|r'_i-r''_i|} \geq x_i \geq 1}} \frac{d^2x_i}{|x_i|^2}~
+ O\Big( \frac{|r'_i|}{R_i},\frac{|r''_i|}{R_i}\Big).
\ee

By applying the symmetrization on eq.(\ref{int}) we get the following
expression for $\C{M}_n$

\ba \label{Tint}
\C{M}_n (U_1,\cdots,U_n) &=& \int_{\scr{4U_1 \geq x_1}} \frac{d^2x_1}{2\pi
|x_1|^2} \cdots \int_{\scr{4U_n \geq x_n}} \frac{d^2x_n}{2\pi |x_n|^2}  \nonu
\\
& &~~~~~~~~~~~~\prod_{p,q=1 \atop p<q}^n
\frac{2|x_p|^2|x_q|^2}{|x_p-x_q|^2|1-x_px_q|^2} F_n\Big(
\frac{(x_1+1)^2}{4x_1}, \cdots, \frac{(x_n+1)^2}{4x_n}\Big),
\ea

\nt where we ignore the irrelevant higher order corrections because of the
condition (\ref{nejat}). The symmetrized integrals are much simpler than the
integrals we started with. Their integrands are rational functions, and they
can be calculated by standard methods. For explicit calculation to the third
order see appendix B.
\vskip .75cm
\pagebreak[3]


\section{Spin-Spin-Energy Density Correlation Function Off-criticality}
\setcounter{equation}{0}
\vskip .25cm
\nopagebreak
\indent

In this section we demonstrate the applicability of our method for the
calculation of more general correlation functions. As a typical example we
calculate the short-distance expansion of the spin-spin-energy density
correlation function to first order. Our starting point is equation
(\ref{OPEn}) for $n=3$

\be \label{sse0}
\la \e(r') \s(r) \s(0) \ra_m = {C_{\s\s m}}^b(r',r;m)\la \fk_b \ra_m,
\ee

\nt where in the sum over $b$ it is sufficient to include scalar composite
operators, which are not total derivatives. Because of the operator algebra of
the critical IM, $\fk_b \in \{[\id],[\e]\}$.
\vskip .5cm
\pagebreak[2]


\nt \underline{Zeroth Order: The Scaling Behavior}
\indent

At $m=0$, due to equations (\ref{id}) and (\ref{sse}), the OPE (\ref{sse0})
becomes

\be \label{Cssm0}
C_{\s\s m}{}^\id(r,r') = \frac{r^{\scriptstyle \frac{3}{4}}}{4 \pi |r'-r||r'|}.
\ee
\vskip .5cm
\pagebreak[2]


\nt \underline{First Order Corrections to Scaling}
\indent

The first order variational formula for general OPE coefficients (\ref{VFcg})
can also be truncated.  The resulting variational formula can be brought into a
more convenient form in which the main term is fully connected. This is
motivated by the results of section 4; the necessary UV subtractions are
separated out of the main term, and therefore our integration methods are
applicable. We can make the main term fully connected by adding and subtracting
an appropriate term. Therefore, using the short-hand notation for truncated
VFs, we have

\be
\pr{m} C_{\s\s m}{}^\id (r,r')  = \pr{m} \la \e(r') \s(r) \s(0) \ra^c + \Big(
\la \s(r)\s(0) \ra - C_{\s\s m}{}^m (r,r') \Big) \pr{m} \la \e \ra,
\ee

\nt which according to the calculational scheme of section 3 is equivalent to

\be \label{OPEexx1}
-\pr{m}C_{\s\s m}{}^{\id}(r,r')= \had{R''}{\infty} \bigg{\{}
r^{-\frac{1}{4}}~\C{N}_1\Big( \frac{r'}{r};\frac{R''}{r}\Big) -\Big(C_{\s\s
m}{}^m(r,r') - r^{-\frac{1}{4}} \Big) \Big( \C{S}_{1,\e}(\frac{R''}{r})
+c_{mm}{}^\id \Big) \bigg{\}},
\ee

\nt where $\C{S}_{1,\e}$ is defined in (\ref{s1e}) and

\ba
\C{N}_1(u';U'')&\!\!\!\equiv& \!\!\!\ant{\scr{U''\geq u''}}{u''}
r^{\frac{9}{4}}~\la \e(ru'') \e(ru') \s(r) \s(0) \ra^c \nonu \\
&\!\!\!=&\!\!\! \frac{1}{4 \pi |u'||u'-1|} ~\int_{\scr{4U''\geq u''}}
\frac{d^2u''}{4 \pi |u''||u''-1|} \bigg[
\frac{|u'+u''-2u'u''|^2-4|u'||u'-1||u''||u''-1|}{|u'-u''|^2} \bigg].~~~~~~
\ea

After the symmetrization transformation for infinitely large $U''$ we obtain

\be \label{nsadeh}
\C{N}_1 \Big( \frac{(x+1)^2}{4x};U''\Big) = \frac{4|x|^2}{\pi |x^2-1|^2}
{}~\int_{\scr{4U'' \geq y \geq 1}} \frac{d^2y}{4\pi|y|^2}~Re \bigg[
\frac{(x-y)(1-\bar{x}\bar{y})}{(\bar{x}-\bar{y})(1-x y)} \bigg],
\ee

\nt This integral can be calculated with standard techniques. With our previous
choice of finite counterterm $c_{mm}{}^\id$, the final results are

\be
C_{\s\s m}{}^m (r',r) = \frac{|2r'-r|^2}{4|r'||r'-r|r^{\scriptstyle
\frac{1}{4}}},
\ee
\vspace{-.35cm}
\be
\pr{m} C_{\s\s m}{}^\id (r',r) = \frac{1}{2\pi r^{\scriptstyle \frac{1}{4}}}
\lB \Big( \frac{|2r'-r|^2}{4|r'||r'-r|} + 1 \Big) \ln r - \ln (16|r'||r'-r|) +
2 \frac{|x|^4 - 1}{|x^2-1|^2} \ln {|x|}\rB,
\ee

\nt where $x=\frac{\displaystyle{1}}{\displaystyle{r}} (2r'-r \pm 2
\sqrt{r'(r'-r)})$. Now inserting back into (\ref{sse0}) we get

\ba \label{EssO1}
\lefteqn{\la \e(r') \s(r) \s(0) \ra_m = \frac{r^{\scriptstyle \frac{3}{4}}}{4
\pi |r'-r||r'|} \Big[ 1 + \frac{|2r'-r|^2 m}{2r} \ln (|mr|e^{\g}/8)} \nonu\\
& &~~~~~~~~~~~~~~~~~~~~~~~~~~~ -\frac{2|r'-r||r'|m}{r} \ln \Big(
\frac{16|r'-r||r'|}{r}\Big) + \frac{mr}{4}\Big( |x|^2-\frac{1}{|x|^2}\Big) \ln
{|x|} + O(m^2r^2) \Big].~~~~~~
\ea

\nt  The two different short-distance limits of this result can be easily
checked to be consistent with those obtained from two-point functions already
calculated. This is a typical example of the applicability of our calculational
scheme for non-trivial higher point functions.
\vskip .75cm
\pagebreak[3]


\section{Conclusion and Future Directions}
\vspace{.25cm}
\nopagebreak
\indent

Three things were done in this paper: First we reviewed the new formalism for
the study of the short-distance structure of renormalizable field theories.
This formalism provided us with the first systematic scheme for perturbative
calculations near a {\em non-trivial\/} fixed point. This review was done both
in general and for the special case of the two dimensional massive Ising model
at zero external magnetic field. In particular, we presented a complete
renormalization group analysis of all the one-point functions of the theory
which contain all non-analytic dependence on the mass of any correlation
function in the theory. Second, we provided the details of the proof of
finiteness of the variational formula for operator product coefficients to all
orders. This amounted to the consistency of the variational formula with the
analyticity of the OPE coefficients with respect to parameters of the theory.
As a by-product, we developed a systematic scheme for perturbative calculations
of the OPE coefficients in the vicinity of any fixed point. Our first main
result was the {\em truncated} and {\em manifestly IR finite} VF
(\ref{tVFOPEo2}), in which there were only a finite number of terms. This VF
can be extended to higher orders readily. Third, we implemented this scheme in
the analysis of the short-distance expansion of the Ising model correlation
functions which only involve two spin operators. The truncated VF for the case
of spin-spin correlation function to $n^{\underline{th}}$ order was given by
equation (\ref{VFn}). In this case we developed a systematic method for
calculating the necessary integrals. Our second main result was the
symmetrization transformation (\ref{symm}) which is at the heart of our method
and transforms these extremely hard integrals into a managable form. In
particular we calculated the spin-spin and spin-spin-energy density correlation
functions to third and first orders, equations (\ref{2spin}) and (\ref{EssO1})
respectively. Our result for the spin-spin correlation function is in complete
agreement with the scaling limit of the well-known exact calculation on the
lattice.

This calculational scheme is formulated in coordinate space and its main
advantage over conventional schemes is its applicability to perturbative
analysis of renormalizable field theories near a non-trivial fixed point. The
fundamental strength of this formalism relies on a natural separation of
infrared and ultraviolet divergences by incorporating OPE from the outset. An
important subtlety to bear in mind is that despite the non-perturbative
validity of this formalism, its applicability to perturbative calculations is
much more intricate for theories with some ill-behaved conjugate operators at
the fixed point\ct{Son4}. However, our scheme is well-suited for a large class
of two dimensional field theories, defined as some perturbation of unitary
minimal models\ct{BPZ,Zam1,Zam2}. One of the difficult steps in the analysis of
these models is to calculate non-trivial {\em finite} integrals of the critical
correlation functions in each theory. We hope that the possibility of
calculating similar integrals in the Ising model sheds some light on the
structure of integrals in the more complicated minimal models. Clearly, there
is much interesting work to be done.

Among the possible future directions for extending this formalism, three stand
out. The first is the renormalization group analysis and perturbative
calculation of the correlation functions in the Ising model with non-zero
external magnetic field. The second is to understand the structure of the flow
between tri-critical and critical Ising models. The third is the completion of
the final step of our program for the Ising model, as outlined in section 3. It
would be extremely instructive to explicitly see that the validity of our
scheme to any order fixes all the anomalous dimensions and the finite
counterterms. It should also clarify the role of integrability of the massive
Ising model in organizing our field-theoretic calculations effectively\ct{MZ2}.
\vskip .5cm

\nt \underline{Acknowledgement}
\indent

We would like to thank H. Sonoda for suggesting the problem, valuable comments,
constant encouragement, and a careful reading of the manuscript.
\vskip .75cm
\pagebreak[3]


\appendix
\setcounter{equation}{0}
\section{The Critical Ising Model}
\vspace{.25cm}

\underline{Operator Content}
\indent

At the critical point the Ising model acquires conformal symmetry, and we
therefore analyze the operator content of the model using the techniques of
conformal field theory (Consult \ct{BPZ} for more details.). The local
operators of the critical Ising model fall into three classes: class $\idf$,
$\ef$, and $\spf$. Each class is denoted by their respective primary operators.
We call an operator primary if

\be \label{primary}
L_n \fk = \B{L}_n \fk = 0 ~~~~~~~ {\rm for} ~~ n>0,
\ee

\nt where $L_n,\B{L}_n(n\in {\bf Z})$ are left- and right-moving Virasoro
operators. In the class of the primary operator $\fk$ a general operator is
given in the following form

\be \label{descendent}
{\fk}^{\{n\}\{\B{n}\}} \equiv L_{-n_1}\cdots L_{-n_N}
\B{L}_{-\B{n}_{\B{1}}}\cdots \B{L}_{-\B{n}_{\B{N}}} \fk.
\ee

\nt To find the linear relation among various operators we need the Virasoro
algebra:

\ba \label{vir}
[ L_m,L_n ] &=& (m-n)L_{m+n} + \frac{1}{24}m(m^2-1)\dd_{m+n,0}, \nonumber \\
{}~[ {\B{L}}_m,{\B{L}}_n ] &=& (m-n){\B{L}}_{m+n} + \frac{1}{24}m(m^2-1)
\dd_{m+n,0}.
\ea

The scale dimension $x$ and the spin $h$ of the local operator
${\fk}^{\{n\}\{\B{n}\}}$, are given by:

\ba
x &=& (\db + n_1 + \cdots + n_N) + (\B{\db} + \B{n}_{\B{1}} + \cdots +
\B{n}_{\B{N}}), \nonumber \\
h &=& (\db + n_1 + \cdots + n_N) - (\B{\db} + \B{n}_{\B{1}} + \cdots +
\B{n}_{\B{N}}),
\ea

\nt where $(\db,\B{\db})$ are the eigenvalues of $(L_0,\B{L}_0)$ for $\fk$; for
$\id$, $\C{E}$, and $\s$ they are given by (0,0), $(\hf,\hf)$, and $(\hh,\hh)$
respectively. Therefore, for scalar operators a lighter notation than the one
in (\ref{descendent}) can be adopted; ${\fk}^{\{n\}}\equiv {\fk}^{\{n\}\{n\}}$.

The Ward identity associated with $SL(2,{\bf C})$ invariance of conformal field
theories implies

\be \label{dif}
L_{-1} = \partial_z \equiv (\partial_x - i \partial_y)/2.
\ee

\nt Therefore, using (\ref{descendent}) and (\ref{vir}) we find that, up to
scale dimension 7, we have the following scalar operators

\be
\C{O}_0 \equiv \id, ~~\C{O}_m \equiv \C{E}, ~~\C{T} \equiv \id^{\{2\}} =
L_{-2}\B{L}_{-2} \id,
\ee

\nt in the families $\idf$ and $\ef$.

The non-trivial operator product expansions between members of these conformal
classes are given by:

\be \label{IsOPE}
\spf\cdot\spf = \idf + \ef, ~~~~~\spf\cdot\ef = \spf, ~~~~~\ef\cdot\ef = \idf.
\ee

\nt The off-critical Ising model is obtained by adding local scalar operators
to the critical hamiltonian density. Therefore the most general hamiltonian for
the off-critical Ising model is

\be \label{IsH}
\C{H}{\{\mu_{\scriptstyle \idf},\mu_{\scriptstyle \ef},\mu_{\scriptstyle
\spf}\}} = \C{H}^\ast + \mu_{\scriptstyle \idf}\int \C{O}_{\scriptstyle \idf} +
\mu_{\scriptstyle \ef}\int \C{O}_{\scriptstyle \ef} + \mu_{\scriptstyle
\spf}\int \C{O}_{\scriptstyle \spf}
\ee

\nt where the term $\mu_{[i]}\C{O}_{[i]}$ denotes a linear combination of all
scalar operators in class $[i]$. By virtue of (\ref{IsOPE}) and the fact that
RG transformations only depend on local physics we can set $\mu_{\spf} = 0$.
Note that, since the OPE of two operators in class $\spf$ contains no operators
in that class, it make sense to speak of off-critical correlation functions
with an even number of spin operators.
\vskip .5cm
\pagebreak[2]


\nt \underline{Correlation Functions}

We start by fixing the normalization of our operators at the fixed point

\be \label{norme}
\la\C{E}(r)\C{E}(0)\ra = \frac{1}{4 \pi^2 r^{2x_{\C{E}}}}
\ee
\be \label{norms}
\la\s(r)\s(0)\ra = \frac{1}{r^{2x_{\s}}}.
\ee

\nt The convention that correlation functions have no subscript means that we
are calculating fixed point quantities.

We now use the results of \ct{DSZ} to calculate all the necessary correlation
functions of the critical Ising model. All the correlation functions are
expressed in terms of the correlators of a free bosonic field $\fk$ whose two
point function is

\be
\la\fk(r)\fk(r')\ra = -\hf\ln|r-r'|,
\ee

\nt which implies

\be
\la e^{im\fk(r)} e^{in\fk(r')} \ra = |r-r'|^{\kk{mn}{2}},
\ee

\nt and the neutrality condition

\be \label{neut}
\la \prod_{i=1}^{n} e^{im_{i}\fk(r_i)} \ra \neq 0  ~~~~~ {\rm iff} ~~~
\sum_{i=1}^{n} m_{i} = 0.
\ee

The non-trivial three point function is given by (See equation (2.27) of ref.
\ct{DSZ})\footnote{Note that our particular choice of normalization in
(\ref{norme}) modifies the equations from ref.\ct{DSZ} by factors of $2 \pi$.}

\be \label{sse}
\la \C{E}(r')\s(r)\s(0) \ra_0 = \frac{1}{\pi} \la \s(r)\s(0) \ra \cdot \la \cos
\fk(r') \cos \fk (r) \cos \fk (0) \ra = \frac{r^{\scriptstyle \frac{3}{4}}}{4
\pi |r'-r||r'|}.
\ee

\nt By letting $r \rightarrow 0$ and using (\ref{norme}) and (\ref{norms}) we
get:

\be \label{Hsse0}
C_{\s\s}{}^m (r;0) = \pi r^{\frac{3}{4}}
\ee

The four point functions of interest are $\la \C{E}\C{E}\s\s \ra$ and $\la
\C{E}\C{E}\C{E}\C{E} \ra$. Similar to (\ref{sse}) we start with (See equation
(2.28) of ref. \ct{DSZ})

\ba \label{228}
\la \C{E}(r'')\C{E}(r')\s(r)\s(0) \ra \cdot \la \s(r)\s(0) \ra + \la
\C{E}(r'')\s(r)\s(0) \ra \cdot \la \C{E}(r')\s(r)\s(0) \ra \nonumber \\
= \frac{1}{\pi^2} \la \cos 2\fk(r'') \cos 2\fk(r')\cos \fk(r) \cos \fk(0) \ra
\ea

\nt and using (\ref{sse}), (\ref{norms}), and (\ref{neut}) we have:

\be \label{esse}
\la \C{E}(r'')\C{E}(r')\s(r)\s(0) \ra = \frac{|rr'' + rr' - 2r''r'|^2}{16 \pi^2
|r''||r-r''||r'||r-r'||r''-r'|^2 r^{\scriptscriptstyle {1 \over 4}}}\cdot
\ee

\nt from ref. \ct{DSZ} we also have:

\be \label{eeee}
\la \C{E}(r_3)\C{E}(r_2)\C{E}(r_1)\C{E}(0) \ra = \sum_{{\rm cyclic
{}~permutation} \atop {\rm ~of} ~(r_3,r_2,r_1)} \frac{1}{(2\pi)^4}
\bigg[\frac{|r_1|^2|r_2-r_3|^2}{|r_2|^2|r_1-r_3|^2|r_3|^2|r_1-r_2|^2} -
\frac{1}{|r_1|^2|r_2-r_3|^2} \bigg].
\ee

The last correlation function we need is $\la \C{E}\C{E}\C{E}\s\s \ra$. As
pointed out in \ct{DSZ}, a slight extension of (\ref{sse}) and (\ref{228})
allows for calculation of correlation function of any even number of spin
operators and an arbitrary number of energy density operators. we therefore
have

\ba
\la \C{E}(r_3)\C{E}(r_2)\C{E}(r_1)\s(r)\s(0) \ra \cdot \la \s(r)\s(0) \ra + \la
\C{E}(r_3)\C{E}(r_2)\s(r)\s(0) \ra \cdot \la \C{E}(r_1)\s(r)\s(0) \ra \nonu\\
+ \la \C{E}(r_3)\C{E}(r_1)\s(r)\s(0) \ra \cdot \la \C{E}(r_2)\s(r)\s(0) \ra
+\la \C{E}(r_2)\C{E}(r_1)\s(r)\s(0) \ra \cdot \la \C{E}(r_3)\s(r)\s(0) \ra
\nonumber\\
= 8 \la \cos 2\fk(r_3)\cos 2\fk(r_2)\cos 2\fk(r_1)\cos \fk(r)\cos \fk(0)  \ra
\ea

\nt from which, using (\ref{esse}), (\ref{sse}), (\ref{norms}), and
(\ref{neut}), we get:

\ba \label{eesse}
\la \C{E}(r_3)\C{E}(r_2)\C{E}(r_1)\s(r)\s(0) \ra &=& \frac{r^{\scriptstyle
\frac{3}{4}}}{16 \pi^3 |r_1||r_1-r||r_2||r_2-r||r_3||r_3-r|}\nonumber\\
& & \times \sum_{{\rm cyclic ~permutation} \atop {\rm ~of} ~(r_3,r_2,r_1)}
\bigg[ \frac{|r_1-r_2|^2|r_3-r|^2|r_3|^2}{|r_3-r_1|^2|r_3-r_2|^2}
- \frac{|rr_1+rr_2-2r_1r_2|^2}{4|r_1-r_2|^2} \bigg].~~~~~~~~
\ea

\nt More general correlation functions can be calculated similarly.
\vskip .75cm
\pagebreak[3]


\setcounter{equation}{0}
\section{Calculation of Spin-Spin Correlation Function to Third Order}

In this appendix we present the details of the calculations of ref. \ct{MZ1}.
This is an application of the analysis of the $n^{\underline{th}}$ corrections
to scaling for the spin-spin correlation function (section 5).
\vskip .5cm
\pagebreak[2]


\nt \underline{Zero-th Order: The Scaling Behavior}
\nopagebreak
\indent

As a consequence of (\ref{id}) and (\ref{norms}) we have

\be
C_{\s\s}{}^{\id}(r)=\frac{1}{r^{1/4}}.
\ee
\vskip .5cm
\pagebreak[2]


\nt \underline{Corrections to Scaling}
\nopagebreak
\indent

For $n=1,2,3$, the equation (\ref{VFn}) reduces to

\ba \label{OPExx1}
-\pr{m}C_{\s\s}{}^{\id}(r)&=& \had{R'}{\infty} \bigg{\{}
r^{\frac{3}{4}}~\C{M}_1\Big( \frac{R'}{r}\Big)- C_{\s\s}{}^m(r) \Big[
\C{S}_{1,\e}\Big( \frac{R'}{r}\Big) +c_{mm}{}^\id \Big] \bigg{\}},
\ea
\be \label{OPEt2e}
\pr{m}^2 C_{\s\s}{}^{\id}(r) = \had{R'}{\infty} \had{R''}{\infty} \bigg{\{ }
r^{\frac{7}{4}} ~\C{M}_2 \Big( \frac{R'}{r}, \frac{R''}{r}\Big) + \pr{m}
C_{\s\s}{}^m(r) \bigg[ \C{S}_{1,\e}\Big( \frac{R'}{r}\Big) + \C{S}_{1,\e}\Big(
\frac{R''}{r}\Big) + 2 {c_{mm}}^\id \bigg] \bigg{\} },
\ee

\nt and

\ba
-\pr{m}^3 C_{\s\s}{}^{\id}(r)&\!\!\!=&\!\!\! \had{R_1}{\infty}
\had{R_2}{\infty} \had{R_3}{\infty} \bigg{\{ } r^{\frac{11}{4}}~ \C{M}_3\Big(
\frac{R_1}{r}, \frac{R_2}{r}, \frac{R_3}{r}\Big) \nonu \\
& &- r^2 C_{\s\s}{}^m(r) \sum_{{\rm cyclic~perm.}} \C{S}_{3,\e} \Big(
\frac{R_1}{r}, \frac{R_2}{r}, \frac{R_3}{r}\Big) - \pr{m}^2 C_{\s\s}{}^m(r)
\sum_{{\rm cyclic~perm.}} \C{S}_{1,\e}\Big( \frac{R_1}{r}\Big)
\bigg{\}},~~~~~~~
\ea

\nt where the expression for $\C{M}_n, ~(n=1,2,3)$ is given by (\ref{main}) and
the subtractions are as follows

\be \label{s1e}
\C{S}_{1,\e}(U_1) = \had{\ep}{0} \Big( \int_{\scr{U_1\geq u_1\geq \frac{\ep}{r}
}} \frac{d^2u_1}{4 \pi^2 |u_1|^2} +\frac{1}{2 \pi} \ln (\ep) \Big),
\ee
\be
\C{S}_{3,\e}(U_1,U_2,U_3) = \int_{\scr{U_1\geq u_1}} \frac{d^2u_1}{2 \pi
|u_1|^2} \int_{\scr{U_2\geq u_2}} \frac{d^2u_2}{2 \pi |u_2|^2}
\int_{\scr{U_3\geq u_3}} \frac{d^2u_3}{2 \pi |u_3|^2} \Big[
\frac{|u_1|^4|u_2-u_3|^2}{2\pi|u_1-u_3|^2|u_1-u_2|^2} -
\frac{|u_2|^2|u_3|^2}{2\pi|u_2-u_3|^2} \Big].
\ee

\nt The explicit expressions for necessary correlation functions are given in
appendix A. For the calculations of the main terms we use the symmetrization
transformation (\ref{symm}). The resulting integrals are

\be
\C{M}_1(U_1)  = \int_{\scr{4U_1 \geq x_1 \geq 1}} \frac{d^2x_1}{4 \pi |x_1|^2},
\ee
\be \label{sadeh2}
\C{M}_2 (U_1,U_2)=  \int_{\scr{4U_1 \geq x_1 \geq 1}}\frac{d^2x_1}{4\pi|x_1|^2}
{}~\int_{\scr{4U_2 \geq x_2 \geq 1}} \frac{d^2x_2}{4\pi|x_2|^2}~Re \bigg[
\frac{(x_1-x_2)(1-\bar{x}_1\bar{x}_2)}{(\bar{x}_1-\bar{x}_2)(1-x_1 x_2)}
\bigg],
\ee
\ba \label{Tint3}
\lefteqn{\C{M}_3(U_1,U_2,U_3) =  \int_{\scr{4U_1 \geq x_1}} \frac{d^2x_1}{2\pi
|x_1|^2} \int_{\scr{4U_2 \geq x_2}} \frac{d^2x_2}{2\pi |x_2|^2}\int_{\scr{4U_3
\geq x_3}} \frac{d^2x_3}{2\pi |x_3|^2}\prod_{p,q=1 \atop p<q}^3
\frac{2|x_p|^2|x_q|^2}{|x_p-x_q|^2|1-x_px_q|^2}} \nonu \\
& &~~~~~~~~~~~~~~~\times \sum_{{\rm cyclic \atop perm.}} \bigg{\{}
|x_1-x_2|^4|1-x_1 x_2|^4|x_3-1|^4 - |x_1-x_3|^2|x_2-x_3|^2|1-x_1 x_3|^2|1-x_2
x_3|^2 \nonu\\
& &~~~~~~~~~~~~~~~~~~~~~~~~~~~~~~~~~~~~~~~~~~~~~~\times \Big(
|x_1^2-1|^2|x_2^2-1|^2 + 2 Re \Big( (x_1-x_2)^2(1-\bar{x}_1\bar{x}_2)^2 \Big)
\Big) \bigg{\}}.~~~~~~
\ea

These integrals do not have any square roots and they are easily calculable.
These allow for the calculation of corrections to the scaling behavior of the
OPE coefficients. The results are

\ba \label{aha}
C_{\s\s}{}^m(r)&=& \pi r^{\frac{3}{4}}, \\
\label{ahaa}
\pr{m}C_{\s\s}{}^\id (r)&=& \frac{1}{r^{\frac{3}{4}}} \ln r, \\
\label{zero}
\pr{m} C_{\s\s}{}^m(r)&=&0, \\ \label{eight}
\pr{m}^2 C_{\s\s}{}^\id (r)&=& \frac{r^{\frac{7}{4}}}{8},\\
\label{sixteen}
\pr{m}^2 C_{\s\s}{}^m (r)&=& \frac{\pi r^{\frac{11}{4}}}{8},\\ \label{third}
\pr{m}^3 C_{\s\s}{}^\id (r)&=& \frac{3r^{\frac{11}{4}}}{16} \ln r.
\ea

\nt The results in equations (\ref{ahaa}) and (\ref{third}) are due to the
particular choice of the finite counterterm $c_{mm}{}^\id$. Here $c_{mm}{}^\id
=\frac{1}{\pi} \ln 2$. Consequently from (\ref{ccm}) we fix the integration
constant $c = \frac{e^\g}{8}$, where $\g$ is the Euler's constant. Recall that
the answer for the correlation functions is independent of the choice of the
finite counterterms (See section 4). With this information  at hand, we arrive
at the following expression for the short-distance expansion of the spin-spin
correlation function

\be \label{2spin}
\la \s(r)\s(0) \ra_m = \frac{1}{r^{\frac{1}{4}}} \Big( 1 + \frac{1}{2} t \ln
(|t|e^{\g}/8) + \frac{1}{16} t^2 + \frac{1}{32} t^3 \ln (|t|e^{\g}/8) +
O(t^4\ln ^2|t|) \Big),
\ee

\nt where $t= mr$. This is in complete agreement with the results of the
scaling limit of the lattice calculation\ct{WMTB}.




\begin{thebibliography}{99}
%
\bibitem{Wil2}
K. ~Wilson and J. Kogut, \prp{12C}(1974) 76.
%
\bibitem{WMTB}
T.T. ~Wu, B.M. ~McCoy, C.A. ~Tracy, and E. ~Barouch, Phys. Rev. {\bf B13}(1976)
316.
%
\bibitem{Zam1}
A.B. ~Zamolodchikov, JETP Lett. {\bf 43}(1986) 730; Sov. J. Phys. {\bf
46}(1988) 1090; \np{358}(1991) 524.
%
\bibitem{Zam2}
A.B. ~Zamolodchikov, Adv. Stud. in Pure Math. {\bf 19}(1989) 641;\\
Rev. in Math. Phys. Vol 1, No 2(1990) 197.
%
\bibitem{Pol1}
A.M. ~Polyakov, JETP Lett. {\bf 12}(1970) 381.\\
J. ~Polchinski, \np{303}(1988) 226.
%
\bibitem{String}
M. ~Green, J. ~Schwartz, and E. ~Witten, {\it String Theory Vol I} (Cambridge
Univ. Press, Cambridge, 1985)\\
A.M. ~Polyakov, {\it Gauge Fields and Strings} (Harwood Academic Publishers
1987)
%
\bibitem{BPZ}
A.A. ~Belavin, A.M. ~Polyakov, and A.B. ~Zamolodchikov, \np{241}(1984) 333.
%
\bibitem{CFT}
{\it Conformal Invariance and Applications to Statistical Mechanics}, eds. C.
{}~Itzykson, H. ~Saleur, and J.B. ~Zuber (World Scientific, Singapore, 1988)
%
\bibitem{flow}
D.A. ~Kastor, E.J. ~Martinec, and S.H. ~ Shenker, \np{316}(1989) 590.\\
P. ~Christe and G. ~Mussardo, \np{330}(1990) 465.\\
M. ~Lassig, G. ~Mussardo, and J.L. ~Cardy, \np{348}(1991) 591.
%
\bibitem{Wil1}
K. ~Wilson, \prv{179}(1969) 1499.
%
\bibitem{Weg1}
F. ~Wegner, Phys. Rev. {\bf B5}(1972) 4529.
%
\bibitem{Weg2}
F. ~Wegner, J. Phys. {\bf A8}(1975) 710.
%
\bibitem{Son1}
H. ~Sonoda, \np{352}(1991) 585.
%
\bibitem{Son2}
H. ~Sonoda, \np{352}(1991) 601.
%
\bibitem{Son3}
H. ~Sonoda, \np{366}(1991) 629.
%
\bibitem{Son4}
H. ~Sonoda, \np{383}(1992) 173.
%
\bibitem{Son5}
H. ~Sonoda, \np{394}(1993) 302.
%
%
\bibitem{Son7}
K. ~Ranganathan, H. ~Sonoda, and B. ~Zwiebach,
{\it Connections on the State-Space over Conformal field Theories}
(hep-th 9304053).
%
\bibitem{Son8}
H. ~Sonoda,
{\it Connections on the Theory Space}
(hep-th 9306119).
%
\bibitem{SML}
T.D. ~Schultz, D.C. ~Mattis, and E.H. ~Lieb, Rev. Mod. Phys. {\bf 36}(1964)
856. %
\bibitem{Par}
G. ~Parisi,
{\it Statistical Field Theory} (Addison-Wesley, Menlo Park, 1988).
%
\bibitem{thooft}
G. ~'t Hooft, \np{62}(1973) 444.
%
\bibitem{Itz}
See, for example, C. ~Itzykson and J.M. ~Drouffe,
{\it Statistical Field Theory Vol. I} (Cambridge Univ. Press, Cambridge, 1989)
%
\bibitem{Kad}
L.P. ~Kadanoff, A.C. ~Brown, Ann. Phys. {\bf 121}(1979) 318.
\bibitem{DSZ}
P. ~Di Francesco, H. ~Saleur, and J.B. ~Zuber \np{290}(1987) 527.
%
\bibitem{MZ1}
B. ~Mikhak and A.M. ~Zarkesh,
{\it The Perturbative Calculation of the Spin-Spin Correlation Function in the
Two Dimensional Ising Model\/}
UCLA Preprint UCLA/93/TEP/48, hep-th/9312202.
%
\bibitem{Dot}
V.I. ~Dotsenko, \np{314}(1989) 687.
%
\bibitem{MZ2}
B. ~Mikhak and A.M. ~Zarkesh, Work in progress.
%
\end{thebibliography}
\end{document}